\documentstyle[graphicx,ulem]{mn2e}

\begin{document}

\title[WiggleZ Survey: BAOs in redshift slices]{The WiggleZ Dark
  Energy Survey: mapping the distance-redshift relation with baryon
  acoustic oscillations}

\author[Blake et al.]{\parbox[t]{\textwidth}{Chris
    Blake$^1$\footnotemark, Eyal A.\ Kazin$^2$, Florian Beutler$^3$,
    Tamara M.\ Davis$^{4,5}$, \\ David Parkinson$^4$, Sarah
    Brough$^6$, Matthew Colless$^6$, Carlos Contreras$^1$, \\ Warrick
    Couch$^1$, Scott Croom$^7$, Darren Croton$^1$, Michael
    J.\ Drinkwater$^4$, \\ Karl Forster$^8$, David Gilbank$^9$, Mike
    Gladders$^{10}$, Karl Glazebrook$^1$, \\ Ben Jelliffe$^7$, Russell
    J.\ Jurek$^{11}$, I-hui Li$^1$, Barry Madore$^{12}$,
    \\ D.\ Christopher Martin$^8$, Kevin Pimbblet$^{13}$, Gregory
    B.\ Poole$^1$, Michael Pracy$^{1,14}$, \\ Rob Sharp$^{6,14}$,
    Emily Wisnioski$^1$, David Woods$^{15}$, Ted K.\ Wyder$^8$ and
    H.K.C. Yee$^{16}$} \\ \\ $^1$ Centre for Astrophysics \&
  Supercomputing, Swinburne University of Technology, P.O. Box 218,
  Hawthorn, VIC 3122, Australia \\ $^2$ Center for Cosmology and
  Particle Physics, New York University, 4 Washington Place, New York,
  NY 10003, United States \\ $^3$ International Centre for Radio
  Astronomy Research, University of Western Australia, 35 Stirling
  Highway, Perth WA 6009, Australia \\ $^4$ School of Mathematics and
  Physics, University of Queensland, Brisbane, QLD 4072, Australia
  \\ $^5$ Dark Cosmology Centre, Niels Bohr Institute, University of
  Copenhagen, Juliane Maries Vej 30, DK-2100 Copenhagen \O, Denmark
  \\ $^6$ Australian Astronomical Observatory, P.O. Box 296, Epping,
  NSW 1710, Australia \\ $^7$ Sydney Institute for Astronomy, School
  of Physics, University of Sydney, NSW 2006, Australia \\ $^8$
  California Institute of Technology, MC 278-17, 1200 East California
  Boulevard, Pasadena, CA 91125, United States \\ $^9$ Astrophysics
  and Gravitation Group, Department of Physics and Astronomy,
  University of Waterloo, Waterloo, ON N2L 3G1, Canada \\ $^{10}$
  Department of Astronomy and Astrophysics, University of Chicago,
  5640 South Ellis Avenue, Chicago, IL 60637, United States \\ $^{11}$
  Australia Telescope National Facility, CSIRO, Epping, NSW 1710,
  Australia \\ $^{12}$ Observatories of the Carnegie Institute of
  Washington, 813 Santa Barbara St., Pasadena, CA 91101, United States
  \\ $^{13}$ School of Physics, Monash University, Clayton, VIC 3800,
  Australia \\ $^{14}$ Research School of Astronomy \& Astrophysics,
  Australian National University, Weston Creek, ACT 2611, Australia
  \\ $^{15}$ Department of Physics \& Astronomy, University of British
  Columbia, 6224 Agricultural Road, Vancouver, BC V6T 1Z1, Canada
  \\ $^{16}$ Department of Astronomy and Astrophysics, University of
  Toronto, 50 St.\ George Street, Toronto, ON M5S 3H4, Canada}

\maketitle

\begin{abstract}
We present measurements of the baryon acoustic peak at redshifts $z =
0.44$, $0.6$ and $0.73$ in the galaxy correlation function of the
final dataset of the WiggleZ Dark Energy Survey.  We combine our
correlation function with lower-redshift measurements from the
6-degree Field Galaxy Survey and Sloan Digital Sky Survey, producing a
stacked survey correlation function in which the statistical
significance of the detection of the baryon acoustic peak is
$4.9$-$\sigma$ relative to a zero-baryon model with no peak.  We fit
cosmological models to this combined baryon acoustic oscillation (BAO)
dataset comprising six distance-redshift data points, and compare the
results to similar fits to the latest compilation of supernovae (SNe)
and Cosmic Microwave Background (CMB) data.  The BAO and SNe datasets
produce consistent measurements of the equation-of-state $w$ of dark
energy, when separately combined with the CMB, providing a powerful
check for systematic errors in either of these distance probes.
Combining all datasets we determine $w = -1.03 \pm 0.08$ for a flat
Universe, consistent with a cosmological constant model.  Assuming
dark energy is a cosmological constant and varying the spatial
curvature, we find $\Omega_{\rm k} = -0.004 \pm 0.006$.
\end{abstract}

\section{Introduction}
\renewcommand{\thefootnote}{\fnsymbol{footnote}}
\setcounter{footnote}{1}
\footnotetext{E-mail: cblake@astro.swin.edu.au}

Measurements of the cosmic distance-redshift relation have always
constituted one of the most important probes of the cosmological
model.  Eighty years ago such observations provided evidence that the
Universe is expanding; more recently they have convincingly suggested
that this expansion rate is accelerating.  The distance-redshift
relation depends on the expansion history of the Universe, which is in
turn governed by its physical contents including the properties of the
``dark energy'' which has been hypothesized to be driving the
accelerating expansion.  One of the most important challenges in
contemporary cosmology is to distinguish between the different
possible physical models for dark energy, which include a material or
scalar field smoothly filling the Universe with a negative
equation-of-state, a modification to the laws of gravity at large
cosmic scales, or the effects of inhomogeneity on cosmological
observations.  Cosmological distance measurements provide one of the
crucial observational datasets to help distinguish between these
different models.

One of the most powerful tools for mapping the distance-redshift
relation is Type Ia supernovae (SNe Ia).  About a decade ago,
observations of nearby and distant SNe Ia provided some of the most
compelling evidence that the expansion rate of the Universe is
accelerating (Riess et al.\ 1998, Perlmutter et al.\ 1999), in
agreement with earlier suggestions based on comparisons of the Cosmic
Microwave Background (CMB) and large-scale structure data (Efstathiou,
Sutherland \& Maddox 1990, Krauss \& Turner 1995, Ostriker \&
Steinhardt 1995).  Since then the sample of SNe Ia available for
cosmological analysis has grown impressively due to a series of large
observational projects which has populated the Hubble diagram across a
range of redshifts.  These projects include the Nearby Supernova
Factory (Copin et al.\ 2006), the Center for Astrophysics SN group
(Hicken et al.\ 2009), the Carnegie Supernova Project (Hamuy et
al.\ 2006) and the Palomar Transient Factory (Law et al.\ 2009) at low
redshifts $z < 0.1$; the Sloan Digital Sky Survey (SDSS) supernova
survey (Kessler et al.\ 2009) at low-to-intermediate redshifts $0.1 <
z < 0.3$; the Supernova Legacy Survey (Astier et al.\ 2006) and
ESSENCE (Wood-Vasey et al.\ 2007) projects at intermediate redshifts
$0.3 < z < 1.0$; and observations by the Hubble Space Telescope at
high redshifts $z > 1$ (Riess et al.\ 2004, 2007; Dawson et
al.\ 2009).  These supernovae data have been collected and analyzed in
a homogeneous fashion in the ``Union'' SNe compilations, initially by
Kowalski et al.\ (2008) and most recently by Amanullah et al.\ (2010)
in the ``Union 2'' sample of 557 SNe Ia.

The utility of these supernovae datasets is now limited by known (and
potentially unknown) systematic errors which could bias cosmological
fits if not handled correctly.  These systematics include
redshift-dependent astrophysical effects, such as potential drifts
with redshift in the relations between colour, luminosity and light
curve shape owing to evolving SNe Ia populations, and systematics in
analysis such as the fitting of light curves, photometric zero-points,
K-corrections and Malmquist bias.  Although these systematics have
been treated very thoroughly in recent supernovae analyses, it is
clearly desirable to cross-check the cosmological conclusions with
other probes of the distance-redshift relation.

A very promising and complementary method for mapping the
distance-redshift relation is the measurement of baryon acoustic
oscillations (BAOs) in the large-scale clustering pattern of galaxies,
and their application as a cosmological standard ruler (Eisenstein, Hu
\& Tegmark 1998, Cooray et al.\ 2001, Eisenstein 2003, Blake \&
Glazebrook 2003, Seo \& Eisenstein 2003, Linder 2003, Hu \& Haiman
2003).  BAOs correspond to a preferred length scale imprinted in the
distribution of photons and baryons by the propagation of sound waves
in the relativistic plasma of the early Universe (Peebles \& Yu 1970,
Sunyaev \& Zeldovitch 1970, Bond \& Efstathiou 1984, Holtzman 1989, Hu
\& Sugiyama 1996, Eisenstein \& Hu 1998).  This length scale, which
corresponds to the sound horizon at the baryon drag epoch denoted by
$r_s(z_d)$, may be predicted very accurately by measurements of the
CMB which yield the physical matter and baryon densities that control
the sound speed, expansion rate and recombination time in the early
Universe: the latest determination is $r_s(z_d) = 153.3 \pm 2.0$ Mpc
(Komatsu et al.\ 2009).  In the pattern of late-time galaxy
clustering, BAOs manifest themselves as a small preference for pairs
of galaxies to be separated by $r_s(z_d)$, causing a distinctive
``baryon acoustic peak'' to be imprinted in the 2-point galaxy
correlation function.  The corresponding signature in Fourier space is
a series of decaying oscillations or ``wiggles'' in the galaxy power
spectrum.

Measurement of BAOs has become an important motivation for galaxy
redshift surveys in recent years.  The small amplitude of the baryon
acoustic peak, and the large size of the relevant scales, implies that
cosmic volumes of order 1 Gpc$^3$ must be mapped with of order $10^5$
galaxies to ensure a robust detection (Tegmark 1997, Blake \&
Glazebrook 2003, Glazebrook \& Blake 2005, Blake et al.\ 2006).
Significant detections of BAOs have now been reported by three
independent galaxy surveys, spanning a range of redshifts $z \le 0.6$:
the SDSS, the WiggleZ Dark Energy Survey, and the 6-degree Field
Galaxy Survey (6dFGS).

The most accurate BAO measurements have been obtained by analyzing the
SDSS, particularly the Luminous Red Galaxy (LRG) component.
Eisenstein et al.\ (2005) reported a convincing detection of the
acoustic peak in the 2-point correlation function of the SDSS Third
Data Release (DR3) LRG sample with effective redshift $z=0.35$.
Percival et al.\ (2010) performed a power-spectrum analysis of the
SDSS DR7 dataset, considering both the main and LRG samples, and
measured the distance-redshift relation at both $z=0.2$ and $z=0.35$
with $\sim 3\%$ accuracy in units of the standard ruler scale.  Other
studies of the SDSS LRG sample, producing broadly similar conclusions,
have been undertaken by Hutsi (2006), Percival et al.\ (2007),
S{\'a}nchez et al.\ (2009) and Kazin et al.\ (2010a).  These studies
of SDSS galaxy samples built on hints of BAOs reported by the 2-degree
Field Galaxy Redshift Survey (Percival et al.\ 2001, Cole et
al.\ 2005) and combinations of smaller datasets (Miller et al.\ 2001).
There have also been potential BAO detections in photometric-redshift
catalogues from the SDSS (Blake et al.\ 2007, Padmanabhan et
al.\ 2007, Crocce et al.\ 2011), although the statistical significance
of these measurements currently remains much lower than that which can
be obtained using spectroscopic redshift catalogues.

These BAO detections have recently been supplemented by new
measurements from two different surveys, which have extended the
redshift coverage of the standard-ruler technique.  In the
low-redshift Universe the 6dFGS has reported a BAO detection at
$z=0.1$ (Beutler et al.\ 2011).  This study produced a $\sim 5\%$
measurement of the standard ruler scale and a new determination of the
Hubble constant $H_0$.  At higher redshifts the WiggleZ Survey has
quantified BAOs at $z=0.6$, producing a $\sim 4\%$ measurement of the
baryon acoustic scale (Blake et al.\ 2011).  Taken together, these
different galaxy surveys have demonstrated that BAO standard-ruler
measurements are self-consistent with the standard cosmological model
established from CMB observations, and have yielded new, tighter
constraints on cosmological parameters.

The accuracy with which BAOs may be used to determine the
distance-redshift relation using current surveys is limited by
statistical rather than systematic errors (in contrast to observations
of SNe Ia).  The measurement error in the large-scale correlation
function, which governs how accurately the preferred scale may be
extracted, is determined by the volume of the large-scale structure
mapped and the number density and bias of the galaxy tracers.  There
are indeed potential systematic errors associated with fitting models
to the BAO signature, which are caused by the modulation of the
pattern of linear clustering laid down in the high-redshift Universe
by the non-linear scale-dependent growth of structure, the distortions
apparent when the signal is observed in redshift-space and the bias
with which galaxies trace the network of matter fluctuations.
However, the fact that the BAOs are imprinted on large, linear and
quasi-linear scales of the clustering pattern means that these
non-linear, systematic distortions are amenable to analytical or
numerical modelling and the leading-order effects are well-understood
(Eisenstein, Seo \& White 2007, Crocce \& Scoccimarro 2008, Matsubara
2008, S{\'a}nchez, Baugh \& Angulo 2008, Smith, Scoccimarro \& Sheth
2008, Seo et al.\ 2008, Padmanabhan \& White 2009).  As such, BAOs in
current datasets are believed to provide a robust probe of the
cosmological model, relatively free of systematic error and dominated
by statistical errors.  In this sense they provide a powerful
cross-check of the distance-redshift relation mapped by supernovae.

In this study we report our final analysis of the baryon acoustic peak
from the angle-averaged correlation function of the completed WiggleZ
Survey dataset, in which we present distance-scale measurements as a
function of redshift between $z=0.44$ and $z=0.73$, including a
covariance matrix which may be applied in cosmological parameter fits.
We also present a new measurement of the correlation function of the
SDSS-LRG sample.  We stack the 6dFGS, SDSS-LRG and WiggleZ correlation
functions to produce the highest-significance detection to date of the
baryon acoustic peak in the galaxy clustering pattern.  We perform
cosmological parameter fits to this latest BAO distance dataset, now
comprising data points at six different redshifts.  By comparing these
fits with those performed on the latest compilation of SNe Ia, we
search for systematic disagreements between these two important probes
of the distance-redshift relation.

The structure of our paper is as follows: in Section \ref{secdata} we
summarize the three galaxy spectroscopic redshift survey datasets
which have provided the most significant BAO measurements.  In Section
\ref{secmod} we outline the modelling of the baryon acoustic peak
applied in this study.  In Section \ref{secwigglez} we report the
measurement and analysis of the final WiggleZ Survey correlation
functions in redshift slices, and in Section \ref{seclrg} we present
the new determination of the correlation function of SDSS LRGs.  In
Section \ref{secstack} we construct a stacked galaxy correlation
function from these surveys and analyze the statistical significance
of the BAO detection contained therein.  In Section \ref{seccosmofit}
we perform cosmological parameter fits to various combinations of BAO,
SNe Ia and CMB data, and we list our conclusions in Section
\ref{secconc}.

\section{Datasets}
\label{secdata}

\subsection{The WiggleZ Dark Energy Survey}

The WiggleZ Dark Energy Survey (Drinkwater et al.\ 2010) is a
large-scale galaxy redshift survey of bright emission-line galaxies
which was carried out at the Anglo-Australian Telescope between August
2006 and January 2011 using the AAOmega spectrograph (Saunders et
al.\ 2004, Sharp et al.\ 2006).  Targets were selected via joint
ultraviolet and optical magnitude and colour cuts using input imaging
from the Galaxy Evolution Explorer (GALEX) satellite (Martin et
al.\ 2005), the Sloan Digital Sky Survey (SDSS; York et al.\ 2000) and
the 2nd Red Cluster Sequence (RCS2) Survey (Gilbank et al.\ 2011).
The survey is now complete, comprising of order $200{,}000$ redshifts
and covering of order $800$ deg$^2$ of equatorial sky.  In this study
we analyzed a galaxy sample drawn from our final set of observations,
after cuts to maximize the contiguity of each survey region.  The
sample includes a total of $N = 158{,}741$ galaxies in the redshift
range $0.2 < z < 1.0$.

\subsection{The 6-degree Field Galaxy Survey}

The 6-degree Field Galaxy Survey (6dFGS, Jones et al.\ 2009) is a
combined redshift and peculiar velocity survey covering nearly the
entire southern sky with the exception of a $10^\circ$ band along the
Galactic plane.  Observed galaxies were selected from the 2MASS
Extended Source Catalog (Jarrett et al.\ 2000) and the redshifts were
obtained with the 6-degree Field (6dF) multi-fibre instrument at the
U.K.\ Schmidt Telescope between 2001 and 2006.  The final 6dFGS sample
contains $75{,}117$ galaxies distributed over $\sim 17{,}000$ deg$^2$
with a mean redshift of $z = 0.052$.  The analysis of the baryon
acoustic peak in the 6dFGS (Beutler et al.\ 2011) utilized all
galaxies selected to $K \leq 12.9$.  We provide a summary of this BAO
measurement in Section \ref{sec6df}.

\subsection{The Sloan Digital Sky Survey Luminous Red Galaxy sample}

The SDSS included the largest-volume spectroscopic LRG survey to date
(Eisenstein et al.\ 2001).  The LRGs were selected from the
photometric component of SDSS, which imaged the sky at high Galactic
latitude in five passbands $u,g,r,i$ and $z$ (Fukugita et al.\ 1996,
Gunn et al.\ 1998) using a $2.5$m telescope (Gunn et al.\ 2006).  The
images were processed (Lupton et al.\ 2001, Stoughton et al.\ 2002,
Pier et al.\ 2003, Ivezic et al.\ 2004) and calibrated (Hogg et
al.\ 2001, Smith et al.\ 2002, Tucker et al.\ 2006), allowing
selection of galaxies, quasars (Richards et al.\ 2002) and stars for
follow-up spectroscopy (Eisenstein et al.\ 2001, Strauss et al.\ 2002)
with twin fibre-fed double spectographs.  Targets were assigned to
plug plates according to a tiling algorithm ensuring nearly complete
samples (Blanton et al.\ 2003).

The LRG sample serves as a good tracer of matter because these
galaxies are associated with massive dark matter halos. The high
luminosity of LRGs enables a large volume to be efficiently mapped,
and their spectral uniformity makes them relatively easy to identify.
In this study we analyze similar LRG catalogues to those presented by
Kazin et al.\ (2010a, 2010b)\footnote{These catalogues and the
  associated survey mask are publicly available at {\tt
    http://cosmo.nyu.edu/$\sim$eak306/SDSS-LRG.html}}, to which we
refer the reader for full details of selection and systematics.  In
particular, in this study we focus on the sample DR7-Full, which
corresponds to all LRGs in the redshift range $0.16 < z < 0.44$ and
absolute magnitude range $-23.2 < M_g < -21.2$.  The sky coverage and
redshift distributions of the LRG samples are presented in Figures 1
and 2 of Kazin et al.\ (2010a).  DR7-Full includes $89{,}791$ LRGs
with average redshift $\langle z \rangle = 0.314$, covering total
volume $1.2 \, h^{-3}$ Gpc$^3$ with average number density $8 \times
10^{-5} \, h^3$ Mpc$^{-3}$.

\section{Modelling the baryon acoustic peak}
\label{secmod}

In this Section we summarize the two models we fitted to the new
baryon acoustic peak measurements presented in this study.  These
models describe the quasi-linear effects which cause the acoustic
feature and correlation function shape to deviate from the
linear-theory prediction.  There are two main aspects to model: a
damping of the acoustic peak caused by the displacement of matter due
to bulk flows, and a distortion in the overall shape of the clustering
pattern due to the scale-dependent growth of structure (Eisenstein et
al.\ 2007, Crocce \& Scoccimarro 2008, Matsubara 2008, S{\'a}nchez et
al.\ 2008, Smith et al.\ 2008, Seo et al.\ 2008, Padmanabhan \& White
2009).  Our models are characterized by four variable parameters: the
physical matter density $\Omega_{\rm m} h^2$ (where $\Omega_{\rm m}$
is the matter density relative to the critical density and $h =
H_0/[100 \, {\rm km} \, {\rm s}^{-1} \, {\rm Mpc}^{-1}]$ is the Hubble
parameter), a scale distortion parameter $\alpha$, a physical damping
scale $\sigma_v$, and a normalization factor $b^2$.  The models for
the correlation function $\xi_{\rm model}$ in terms of separation $s$
can be written in the form
\begin{equation}
\xi_{\rm model}(s) = b^2 \, \xi_{\rm fid}(\Omega_{\rm m} h^2,
\sigma_v, \alpha s) .
\label{eqximod}
\end{equation}
The physical matter density $\Omega_{\rm m} h^2$ determines (to first
order) both the overall shape of the matter correlation function and
the length scale of the standard ruler, by determining the physics
before recombination.  The scale distortion parameter $\alpha$ relates
the distance-redshift relation at the effective redshift of the sample
to the fiducial value used to construct the correlation function
measurement, in terms of the $D_V$ parameter (Eisenstein et al.\ 2005,
Padmanabhan \& White 2008, Kazin, S{\'a}nchez \& Blanton 2011):
\begin{equation}
D_V(z_{\rm eff}) = \alpha \, D_{V,{\rm fid}}(z_{\rm eff}) ,
\end{equation}
where $D_V$ is a composite of the physical angular-diameter distance
$D_A(z)$ and Hubble parameter $H(z)$, which respectively govern
tangential and radial separations in a cosmological model:
\begin{equation}
D_V(z) = \left[ (1+z)^2 D_A(z)^2 \frac{cz}{H(z)} \right]^{1/3} .
\end{equation}
The damping scale $\sigma_v$ quantifies the typical displacement of
galaxies from their initial locations in the density field due to bulk
flows, resulting in a ``washing-out'' of the baryon oscillations at
low redshift.  The normalization factor $b^2$, marginalized in our
analysis, models the effects of linear galaxy bias and large-scale
redshift-space distortions.

\subsection{Default correlation function model}

In our first, default, model we constructed the fiducial correlation
function $\xi_{\rm fid}$ in Equation \ref{eqximod} in a similar manner
to Eisenstein et al.\ (2005) and Blake et al.\ (2011).  First, we
generated a linear power spectrum $P_{\rm L}(k)$ as a function of
wavenumber $k$ for a given $\Omega_{\rm m} h^2$ using the CAMB
software package (Lewis, Challinor \& Lasenby 2000).  We fixed the
values of the other cosmological parameters using a fiducial model
consistent with the latest fits to the Cosmic Microwave Background
(Komatsu et al.\ 2011): Hubble parameter $h = 0.71$, physical baryon
density $\Omega_{\rm b} h^2 = 0.0226$, primordial spectral index
$n_{\rm s} = 0.96$ and normalization $\sigma_8 = 0.8$.  We also used
the fitting formulae of Eisenstein \& Hu (1998) to generate a
corresponding ``no-wiggles'' reference spectrum $P_{\rm ref}(k)$,
possessing a similar shape to $P_{\rm L}(k)$ but with the baryon
oscillation component deleted, which we also use in the clustering
model as explained below.

We then incorporated the damping of the baryon acoustic peak caused by
the displacement of matter due to bulk flows (Eisenstein et al.\ 2007,
Crocce \& Scoccimarro 2008, Matsubara 2008) by interpolating between
the linear and reference power spectra using a Gaussian damping term
$g(k) \equiv \exp{(-k^2 \sigma_v^2)}$:
\begin{equation}
P_{\rm damped}(k) = g(k) \, P_{\rm L}(k) + [1 - g(k)] \, P_{\rm ref}(k) .
\label{eqpkdamp}
\end{equation}
The magnitude of the damping coefficient $\sigma_v$ can be estimated
for a given value of $\Omega_{\rm m} h^2$ using the first-order
prediction of perturbation theory (Crocce \& Scoccimarro 2008):
\begin{equation}
\sigma_v^2 = \frac{1}{6\pi^2} \int P_{\rm L}(k) \, dk .
\label{eqsigv}
\end{equation}
However, this relation provides only an approximation to the true
non-linear damping (Taruya et al.\ 2010), and we chose to marginalize
over $\sigma_v$ as a free parameter in our analysis.  We note that
$\sigma_v$ is closely related to the parameter $k_*$ defined by
S{\'a}nchez et al.\ (2008), in the sense that $\sigma_v^2 = 1/2k_*^2$.

We included the boost in small-scale clustering power due to the
non-linear scale-dependent growth of structure using the ``halofit''
prescription of Smith et al.\ (2003), as applied to the no-wiggles
reference spectrum:
\begin{equation}
P_{\rm NL}(k) = \left[ \frac{P_{\rm ref,halofit}(k)}{P_{\rm ref}(k)} \right]
\times P_{\rm damped}(k) .
\end{equation}
Finally, we transformed $P_{\rm NL}(k)$ into the correlation function
appearing in Equation \ref{eqximod}:
\begin{equation}
\xi_{\rm fid}(s) = \frac{1}{2\pi^2} \int dk \, k^2 \, P_{\rm NL}(k)
\left[ \frac{\sin{(ks)}}{ks} \right] .
\label{eqxinl}
\end{equation}

\subsection{Comparison correlation function model}

The second, comparison model we considered for the fiducial
correlation function $\xi_{\rm fid}$ was motivated by perturbation
theory (Crocce \& Scoccimarro 2008, S{\'a}nchez et al.\ 2008):
\begin{equation}
\xi_{\rm fid}(s) = \xi_{\rm L}(s) \otimes \exp{(-s^2/2\sigma_v^2)} + A_{\rm
  MC} \, \frac{d\xi_{\rm L}(s)}{ds} \, \xi_1(s) .
\label{eqxipt}
\end{equation}
In this relation $\xi_{\rm L}(s)$ is the linear model correlation
function corresponding to the linear power spectrum $P_{\rm L}(k)$.
The symbol $\otimes$ denotes convolution by the Gaussian damping
$\sigma_v$, which we evaluated as
\begin{eqnarray}
& & \xi_{\rm L}(s) \otimes \exp{(-s^2/2\sigma_v^2)} \nonumber \\ &=&
  \frac{1}{2\pi^2} \int dk \, k^2 \, P_{\rm L}(k) \, \exp{(-k^2
    \sigma_v^2)} \left[ \frac{\sin{(ks)}}{ks} \right] ,
\end{eqnarray}
and $\xi_1$ is defined by Equation 32 in Crocce \& Scoccimarro (2008):
\begin{equation}
\xi_1(s) = \frac{1}{2\pi^2} \int dk \, k \, P_{\rm L}(k) \, j_1(ks) ,
\end{equation}
where $j_1(x)$ is the spherical Bessel function of order 1.  $A_{\rm
  MC} = 1$ (fixed in our analysis) is a ``mode-coupling'' term that
restores the small-scale shape of the correlation function and causes
a slight shift in the peak position compared to the linear-theory
prediction.  The model of Equation \ref{eqxipt} has been shown to
yield unbiased results in baryon acoustic peak fits by S{\'a}nchez et
al.\ (2008, 2009).

\section{WiggleZ baryon acoustic peak measurements in redshift slices}
\label{secwigglez}

In this Section we describe our measurement and fitting of the baryon
acoustic peak in the WiggleZ Survey galaxy correlation function in
three overlapping redshift ranges: $0.2 < z < 0.6$, $0.4 < z < 0.8$
and $0.6 < z < 1.0$.  Our methodology closely follows that employed by
Blake et al.\ (2011), to which we refer the reader for full details.

\subsection{Correlation function measurements}

\begin{figure*}
\begin{center}
\resizebox{13cm}{!}{\rotatebox{270}{\includegraphics{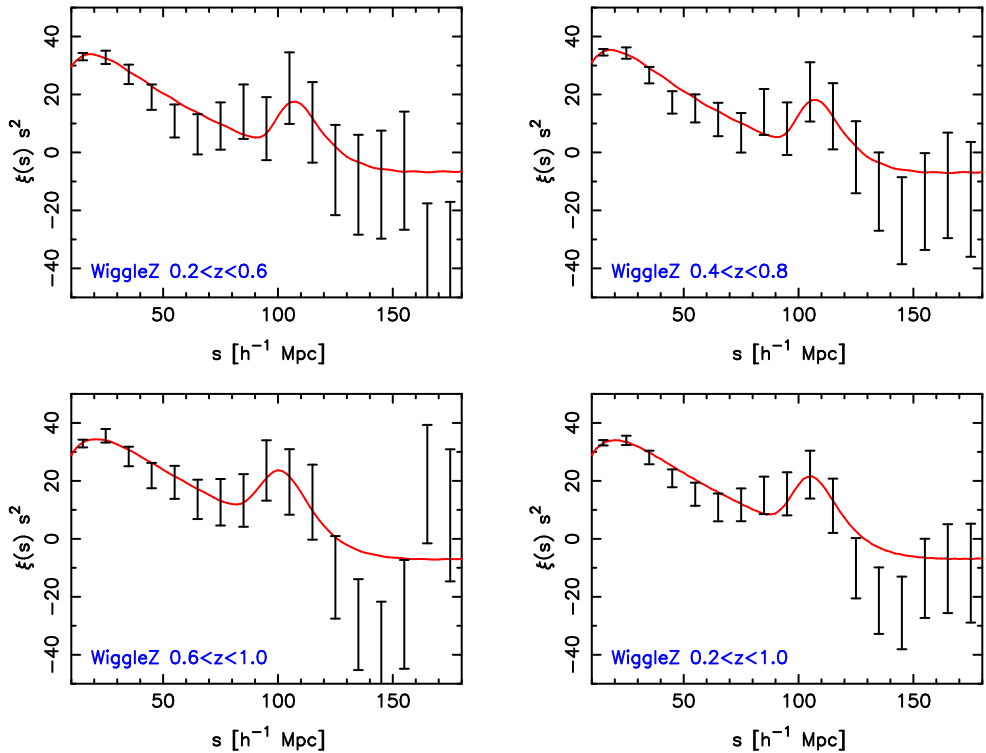}}}
\end{center}
\caption{Measurements of the galaxy correlation function $\xi(s)$,
  combining different WiggleZ survey regions, for the redshift ranges
  $0.2 < z < 0.6$, $0.4 < z < 0.8$, $0.6 < z < 1.0$ and $0.2 < z <
  1.0$, plotted in the combination $s^2 \, \xi(s)$ where $s$ is the
  co-moving redshift-space separation.  The best-fitting clustering
  models in each case, varying the parameters $\Omega_{\rm m} h^2$,
  $\alpha$, $\sigma_v$ and $b^2$ as described in Section \ref{secmod},
  are overplotted as the solid lines.  Significant detections of the
  baryon acoustic peak are obtained in each separate redshift slice.}
\label{figxiwigglez}
\end{figure*}

We measured the angle-averaged 2-point correlation function $\xi(s)$
for each WiggleZ survey region using the Landy-Szalay (1993)
estimator:
\begin{equation}
\xi(s) = \frac{DD(s) - 2 \, DR(s) + RR(s)}{RR(s)} ,
\label{eqxiest}
\end{equation}
where $DD(s)$, $DR(s)$ and $RR(s)$ are the data-data, data-random and
random-random weighted pair counts in separation bin $s$, where each
random catalogue contains the same number of galaxies as the real
dataset.  We assumed a fiducial flat $\Lambda$CDM cosmological model
with matter density $\Omega_{\rm m} = 0.27$ to convert the galaxy
redshifts and angular positions to spatial co-moving co-ordinates.  In
the construction of the pair counts each data or random galaxy $i$ was
assigned a weight $w_i = 1/(1+n_i P_0)$, where $n_i$ is the survey
number density at the location of the $i$th galaxy (determined by
averaging over many random catalogues) and $P_0 = 5000 \, h^{-3}$
Mpc$^3$ is a characteristic power spectrum amplitude at the physical
scales of interest.  The $DR$ and $RR$ pair counts were determined by
averaging over 10 random catalogues, which were constructed using the
selection-function methodology described by Blake et al.\ (2010).  We
measured the correlation function in $10 \, h^{-1}$ Mpc separation
bins in three overlapping redshift slices $0.2 < z < 0.6$, $0.4 < z <
0.8$ and $0.6 < z < 1.0$.  The effective redshift $z_{\rm eff}$ of the
correlation function measurement in each slice was determined as the
weighted mean redshift of the galaxy pairs in the separation bin $100
< s < 110 \, h^{-1}$ Mpc, where the redshift of a pair is simply the
average $(z_1+z_2)/2$, and the weighting is $w_1 w_2$ where $w_i$ is
defined above.  For the three redshift slices in question we obtained
values $z_{\rm eff} = 0.44$, $0.60$ and $0.73$.

We determined the covariance matrix of the correlation function
measurement in each survey region using an ensemble of 400 lognormal
realizations, using the method described by Blake et al.\ (2011).
Lognormal realizations provide a reasonably accurate galaxy clustering
model for the linear and quasi-linear scales which are important for
the modelling of baryon oscillations.  They are more reliable than
jack-knife errors, which provide a poor approximation for the
correlation function variance on BAO scales because the pair
separations of interest are usually comparable to the size of the
jack-knife regions, which are then not strictly independent.  We note
that the lognormal covariance matrix only includes the effects of the
survey window function, and neglects the covariance due to the
non-linear growth of structure and redshift-space effects.  The full
non-linear covariance matrix may be studied with the aid of a large
set of N-body simulations (Rimes \& Hamilton 2005, Takahashi et
al.\ 2011).  Work is in progress to construct such a simulation set
for WiggleZ galaxies, although this is a challenging computational
problem because the typical dark matter haloes hosting the
star-forming galaxies mapped by WiggleZ are $\sim 20$ times lower in
mass than the Luminous Red Galaxy sample described in Section
\ref{seclrg}, requiring high-resolution large-volume simulations.
However, we note that Takahashi et al.\ (2011) demonstrated that the
impact of using the full non-linear covariance matrix on the accuracy
of extraction of baryon acoustic oscillations is small, so we do not
expect our measurements to be compromised significantly through using
lognormal realizations to estimate the covariance matrix.

\begin{figure*}
\begin{center}
\resizebox{13cm}{!}{\rotatebox{270}{\includegraphics{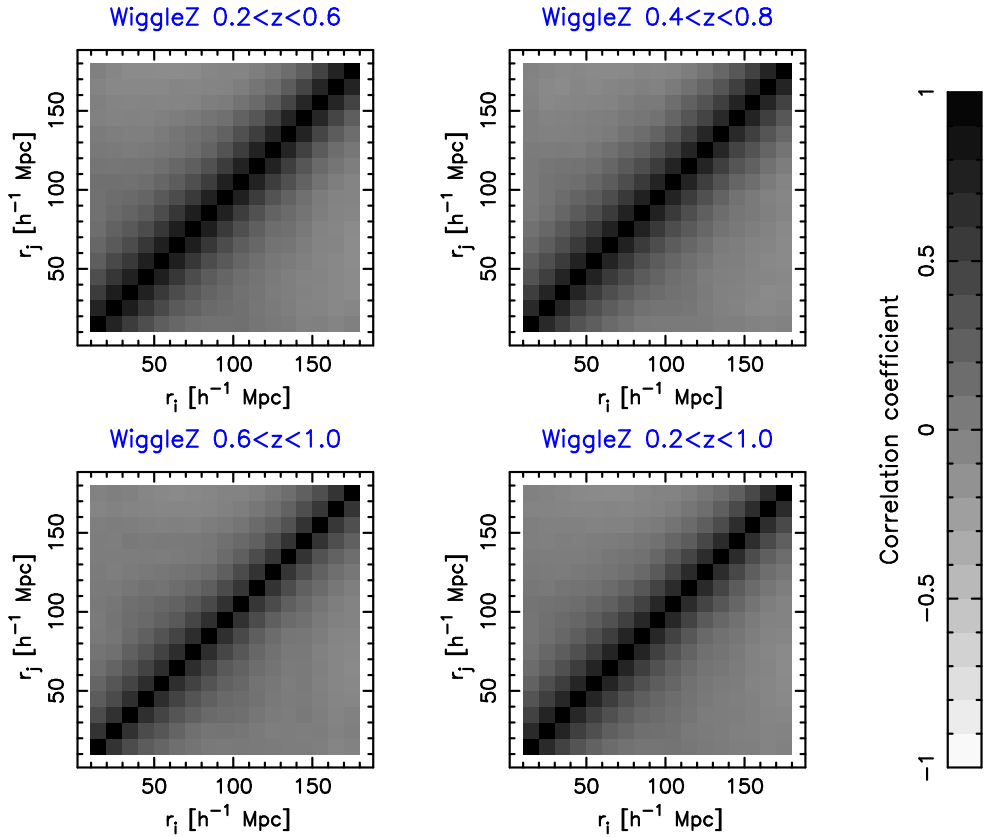}}}
\end{center}
\caption{The amplitude of the cross-correlation $C_{ij}/\sqrt{C_{ii}
    C_{jj}}$ of the covariance matrix $C_{ij}$ for the combined
  WiggleZ correlation function measurements for the redshift ranges
  $0.2 < z < 0.6$, $0.4 < z < 0.8$, $0.6 < z < 1.0$ and $0.2 < z <
  1.0$, determined using lognormal realizations.}
\label{figcovwigglez}
\end{figure*}

We combined the correlation function measurements and corresponding
covariance matrices for the different survey regions using optimal
inverse-variance weighting in each separation bin (see equations 8 and
9 in White et al.\ 2011):
\begin{equation}
\uline{\xi}_{\rm comb} = \uuline{C}_{\rm comb} \sum_{{\rm regions} \,
  n} \uuline{C}_n^{-1} \, \uline{\xi}_n ,
\label{eqxicomb}
\end{equation}
\begin{equation}
\uuline{C}_{\rm comb}^{-1} = \sum_{{\rm regions} \, n} \uuline{C}_n^{-1}
\label{eqcovcomb}
\end{equation}
In these equations, $\uline{\xi}_n$ and $\uline{\xi}_{\rm comb}$ are
vectors representing the correlation function measurements in region
$n$ and the optimally-combined correlation function, and
$\uuline{C}_n$ and $\uuline{C}_{\rm comb}$ are the covariance matrices
corresponding to these two measurement vectors (with inverses
$\uuline{C}_n^{-1}$ and $\uuline{C}_{\rm comb}^{-1}$).  This method
produces an almost identical result to combining the individual pair
counts and then estimating the correlation function using Equation
\ref{eqxiest}.  The combined correlation functions in the three
redshift slices are displayed in Figure \ref{figxiwigglez}, together
with a total WiggleZ correlation function for the whole redshift range
$0.2 < z < 1.0$ which was constructed by combining the separate
measurements for $0.2 < z < 0.6$ and $0.6 < z < 1.0$.  The
corresponding lognormal covariance matrices for each measurement are
shown in Figure \ref{figcovwigglez}.

\subsection{Parameter fits}
\label{secwigfit}

We fitted the first, default correlation function model described in
Section \ref{secmod} to the WiggleZ measurements in redshift slices
$0.2 < z < 0.6$, $0.4 < z < 0.8$ and $0.6 < z < 1.0$, varying
$\Omega_{\rm m} h^2$, $\alpha$, $\sigma_v$ and $b^2$.  Our default
fitting range was $10 < s < 180 \, h^{-1}$ Mpc (following Eisenstein
et al.\ 2005), where $10 \, h^{-1}$ Mpc is an estimate of the minimum
scale of validity for the quasi-linear theory described in Section
\ref{secmod}.  This minimum scale is a quantity which depends on the
survey redshift and galaxy bias (which control the amplitude of the
non-linear, scale-dependent contributions to the shape of the
correlation function) together with the signal-to-noise of the
measurement.  When fitting Equation \ref{eqxinl} to the WiggleZ Survey
correlation function we find no evidence for a systematic variation in
the derived BAO parameters when we vary the minimum fitted scale over
the range $10 \le s_{\rm min} \le 50 \, h^{-1}$ Mpc.

We minimized the $\chi^2$ statistic using the full data covariance
matrix derived from lognormal realizations.  The fitting results,
including the marginalized parameter measurements, are displayed in
Table \ref{tabbaofit}.  The minimum values of $\chi^2$ for the model
fits in the three redshift slices were $11.4$, $10.1$ and $13.7$ for
13 degrees of freedom, indicating that our model provides a good fit
to the data.  The best-fitting scale distortion parameters, which
provide the value of $D_V(z_{\rm eff})$ for each redshift slice, are
all consistent with the fiducial distance-redshift model (a flat
$\Lambda$CDM Universe with $\Omega_{\rm m} = 0.27$) with marginalized
errors of $9.1\%$, $6.5\%$ and $6.4\%$ in the three redshift slices.
The best-fitting matter densities $\Omega_{\rm m} h^2$ are consistent
with the latest analyses of the CMB (Komatsu et al.\ 2011).  The
damping parameters $\sigma_v$ are not well-constrained using our data,
but the allowed range is consistent with the predictions of Equation
\ref{eqsigv} for our fiducial model (which are $\sigma_v = (4.8, 4.5,
4.2) \, h^{-1}$ Mpc for the three redshift slices).  When fitting
$\sigma_v$ we only permit it to vary over the range $\sigma_v \ge 0$.

\begin{figure}
\begin{center}
\resizebox{8cm}{!}{\rotatebox{270}{\includegraphics{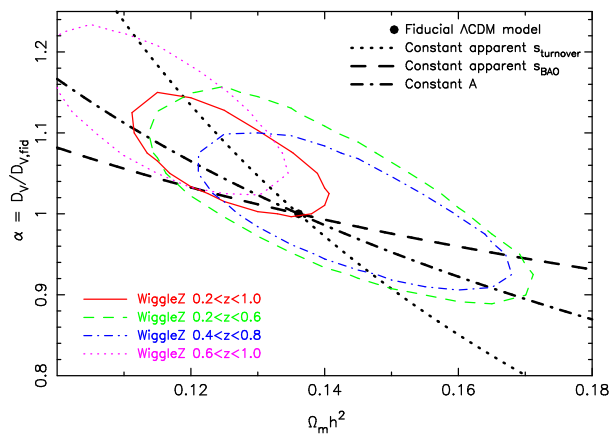}}}
\end{center}
\caption{Probability contours of the physical matter density
  $\Omega_{\rm m} h^2$ and scale distortion parameter $\alpha$
  obtained by fitting to the WiggleZ survey combined correlation
  function in four redshift ranges $0.2 < z < 0.6$, $0.4 < z < 0.8$,
  $0.6 < z < 1.0$ and $0.2 < z < 1.0$.  The heavy dashed and dotted
  lines are the degeneracy directions which are expected to result
  from fits involving respectively just the acoustic peak, and just
  the shape of a pure CDM power spectrum.  The heavy dash-dotted line
  represents a constant value of the acoustic ``A'' parameter defined
  by Equation \ref{eqaz}, which is the parameter best-measured by the
  WiggleZ correlation function data.  The solid circle represents the
  location of our fiducial cosmological model.  The contour level in
  each case encloses regions containing $68.27\%$ of the total
  likelihood.}
\label{figprobwigglez}
\end{figure}

The 2D probability contours for $\Omega_{\rm m} h^2$ and $\alpha$,
marginalizing over $\sigma_v$ and $b^2$, are displayed in Figure
\ref{figprobwigglez}.  The measurement of $\alpha$ (hence $D_V =
\alpha \, D_{V,{\rm fid}}$) is significantly correlated with the
matter density, which controls the shape of the clustering pattern.

We indicate three degeneracy directions in the parameter space of
Figure \ref{figprobwigglez}.  The first direction (the dashed line)
corresponds to a constant measured acoustic peak separation,
i.e.\ $\alpha/r_s(z_d) = {\rm constant}$, where $r_s(z_d)$ is the
sound horizon at the drag epoch as a function of $\Omega_{\rm m} h^2$,
determined using the fitting formula quoted in Equation 12 of Percival
et al.\ (2010).  This parameter degeneracy would be expected in the
case that just the baryon acoustic peak is driving the model fits,
such that the measured low-redshift distance $\alpha \, D_{V,{\rm
    fid}}$ is proportional to the standard ruler scale $r_s(z_d)$.

The second direction (the dotted line) illustrated in Figure
\ref{figprobwigglez} represents the degeneracy resulting from a
constant measured shape of a Cold Dark Matter (CDM) power spectrum,
i.e.\ $\Omega_{\rm m} h^2 \times \alpha = {\rm constant}$.  We note
here the consistency between this scaling and the ``shape parameter''
$\Gamma = \Omega_{\rm m} h$ used to parameterize the CDM transfer
function (Bardeen et al.\ 1986).  This shape parameter assumes that
wavenumbers are observed in units of $h$ Mpc$^{-1}$, but the standard
ruler scale encoded in baryon acoustic oscillations is calibrated by
the CMB in units of Mpc, with no factor of $h$.

The third direction (the dash-dotted line) shown in Figure
\ref{figprobwigglez}, which best describes the degeneracy in our data,
corresponds to a constant value of the acoustic parameter $A(z)$
introduced by Eisenstein et al.\ (2005),
\begin{equation}
A(z) \equiv \frac{100 \, D_V(z) \sqrt{\Omega_{\rm m} h^2}}{c \, z} ,
\label{eqaz}
\end{equation}
which appears in Figure \ref{figprobwigglez} as $\sqrt{\Omega_{\rm m}
  h^2} \times \alpha = {\rm constant}$.  We note that the values of
$A(z)$ predicted by any cosmological model are independent of $h$,
because $D_V$ is proportional to $h^{-1}$.

The acoustic parameter $A(z)$ provides the most appropriate
description of the distance-redshift relation determined by a BAO
measurement in which both the clustering shape and acoustic peak are
contributing toward the fit, such that the whole correlation function
is being used as a standard ruler (Eisenstein et al.\ 2005,
S{\'a}nchez et al.\ 2008, Shoji et al.\ 2009).  In this case, the
resulting measurement of $A(z)$ is approximately uncorrelated with
$\Omega_{\rm m} h^2$.  We repeated our BAO fit to the WiggleZ
correlation functions in redshift slices using the parameter set $(A,
\Omega_{\rm m} h^2, \sigma_v, b^2)$.  The marginalized values of
$A(z)$ we obtained are quoted in Table \ref{tabbaofit}, and correspond
to measurements of the acoustic parameter with accuracies $7.2\%$,
$4.5\%$ and $5.0\%$ in the three redshift slices.

We also fitted our data with the parameter set $(d_z, \Omega_{\rm m}
h^2, \sigma_v, b^2)$, where $d_z \equiv r_s(z_d)/D_V(z)$.  Results are
again listed in Table \ref{tabbaofit}, corresponding to measurements
of $d_z$ with accuracies $7.8\%$, $4.7\%$ and $5.4\%$ in the three
redshift slices.  We note that, unlike for the case of $A(z)$, these
measurements of $d_z$ are correlated with the matter density
$\Omega_{\rm m} h^2$, due to the orientation of the parameter
degeneracy directions in Figure \ref{figprobwigglez} (noting that
constant $d_z$ corresponds to the ``constant measured acoustic peak''
case defined above).

\begin{table*}
\begin{center}
\caption{Results of fitting the four-parameter model $(\Omega_{\rm m}
  h^2, \alpha, \sigma_v, b^2)$ to the WiggleZ correlation functions in
  three redshift slices, together with the results for the full
  sample.  The effective redshifts of the measurement in each slice
  are listed in Column 2, and the corresponding values of $D_V$ for
  the fiducial cosmological model appear in Column 3.  The values of
  $\chi^2$ for the best-fitting models are quoted in Column 4, for 13
  degrees of freedom.  Columns 5, 6 and 7 show the marginalized
  measurements of the matter density parameter $\Omega_{\rm m} h^2$,
  scale distortion parameter $\alpha$ and damping scale $\sigma_v$ in
  each redshift slice.  Corresponding measurements of the BAO
  distilled parameters $A(z)$ and $d_z$ are displayed in Columns 8 and
  9.  The measured values of $D_V$ in each redshift slice are given by
  $\alpha \, D_{V,{\rm fid}}$.}
\label{tabbaofit}
\begin{tabular}{ccccccccc}
\hline
Sample & $z_{\rm eff}$ & $D_{V,{\rm fid}}$ & $\chi^2$ & $\Omega_{\rm m} h^2$ & $\alpha$ & $\sigma_v$ & $A(z_{\rm eff})$ & $d_{z_{\rm eff}}$ \\
& & [Mpc] & & & & [$h^{-1}$ Mpc] & & \\
\hline
WiggleZ - $0.2<z<0.6$ & $0.44$ & $1617.8$ & $11.4$ & $0.143 \pm 0.020$ & $1.024 \pm 0.093$ & $4.5 \pm 3.5$ & $0.474 \pm 0.034$ & $0.0916 \pm 0.0071$ \\
WiggleZ - $0.4<z<0.8$ & $0.60$ & $2085.4$ & $10.1$ & $0.147 \pm 0.016$ & $1.003 \pm 0.065$ & $4.1 \pm 3.4$ & $0.442 \pm 0.020$ & $0.0726 \pm 0.0034$ \\
WiggleZ - $0.6<z<1.0$ & $0.73$ & $2421.9$ & $13.7$ & $0.120 \pm 0.013$ & $1.113 \pm 0.071$ & $4.4 \pm 3.2$ & $0.424 \pm 0.021$ & $0.0592 \pm 0.0032$ \\
WiggleZ - $0.2<z<1.0$ & $0.60$ & $2085.4$ & $11.5$ & $0.127 \pm 0.011$ & $1.071 \pm 0.053$ & $4.4 \pm 3.3$ & $0.441 \pm 0.017$ & $0.0702 \pm 0.0032$ \\
\hline
\end{tabular}
\end{center}
\end{table*}

As a check for systematic modelling errors, we repeated the fits to
the WiggleZ correlation functions using the second acoustic peak model
described in Section \ref{secmod}, motivated by perturbation theory,
fitting the data over the same range of scales.  The marginalized
measurements of $\alpha$ in the three redshift slices were $(1.032 \pm
0.093, 0.981 \pm 0.060, 1.091 \pm 0.079)$, to be compared with the
results for the default model quoted in Table \ref{tabbaofit}.  The
amplitude of the systematic error in the fitted scale distortion
parameter is hence significantly lower than the statistical error in
the measurement (by at least a factor of 3 in all cases).

We assessed the statistical significance of the BAO detections in each
redshift slice by repeating the parameter fits replacing the model
correlation function with one generated using the ``no-wiggles''
reference power spectrum $P_{\rm ref}(k)$ as a function of
$\Omega_{\rm m} h^2$ (Eisenstein \& Hu 1998).  The minimum values
obtained for the $\chi^2$ statistic for the fits in the three redshift
slices were $15.2$, $15.1$ and $19.4$, indicating that the model
containing baryon oscillations was favoured by $\Delta \chi^2 = 3.8$,
$5.0$ and $5.7$ (with the same number of parameters fitted).  These
intervals correspond to detections of the baryon acoustic peaks in the
redshift slices with statistical significances between $1.9$-$\sigma$
and $2.4$-$\sigma$.  We note that the marginalized uncertainty in the
scale distortion parameter for the no-wiggles model fit degrades by a
factor of between two and three compared to the fit to the full model,
demonstrating that the acoustic peak is very important for
establishing the distance constraints from our measurements.

We used the same approach to determine the statistical significance of
the BAO detection in the full WiggleZ redshift span $0.2 < z < 1.0$,
after combining the correlation function measurements in the redshift
slices $0.2 < z < 0.6$ and $0.6 < z < 1.0$.  In this case the model
containing baryon oscillations was favoured by $\Delta \chi^2 = 7.7$,
corresponding to a statistical significance of $2.8$-$\sigma$ for the
detection of the baryon acoustic peak.

\subsection{Covariances between redshift slices}

We used the ensemble of lognormal realizations to quantify the
covariance between the BAO measurements in the three overlapping
WiggleZ redshift slices.  For each of the 400 lognormal realizations
in every WiggleZ region, we measured correlation functions for the
redshift ranges $\Delta z_1 \equiv 0.2 < z < 0.6$, $\Delta z_2 \equiv
0.4 < z < 0.8$ and $\Delta z_3 \equiv 0.6 < z < 1.0$ and combined
these correlation functions for the different regions using
inverse-variance weighting.  We then fitted the default clustering
model described in Section \ref{secmod} to each of the 400 combined
correlation functions for the three redshift slices.

\begin{figure*}
\begin{center}
\resizebox{13cm}{!}{\rotatebox{270}{\includegraphics{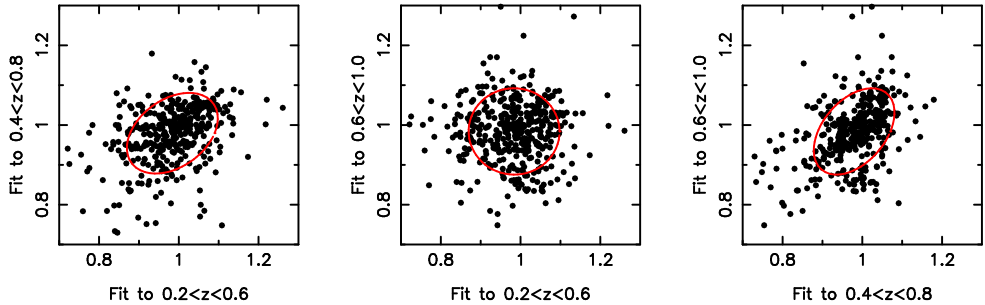}}}
\end{center}
\caption{These panels illustrate the correlations between the scale
  distortion parameters $\alpha$ fitted to correlation functions for
  three overlapping WiggleZ redshift slices using 400 lognormal
  realizations.  The red ellipses represent the derived correlation
  coefficients between these measurements.}
\label{figcovslice}
\end{figure*}

Figure \ref{figcovslice} displays the correlations between the 400
marginalized values of the scale-distortion parameter $\alpha$ for
every pair of redshift slices.  As expected, significant correlations
are found in the values of $\alpha$ obtained from fits to the
overlapping redshift ranges $(\Delta z_1, \Delta z_2)$ and $(\Delta
z_2, \Delta z_3)$, whereas the fits to the non-overlapping pair
$(\Delta z_1, \Delta z_3)$ produce an uncorrelated measurement (within
the statistical noise).  The corresponding correlation coefficients
for the overlapping pairs are $\rho_{12} = 0.369$ and $\rho_{23} =
0.438$, where $\rho_{ij} \equiv C_{ij}/\sqrt{C_{ii} C_{jj}}$ in terms
of the covariances $C_{ij} \equiv \langle \alpha_i \alpha_j \rangle -
\langle \alpha_i \rangle \langle \alpha_j \rangle$.  Table
\ref{tabwigcov} contains the resulting inverse covariance matrix for
the measurements of $A(z)$ in the three redshift slices, that should
be used in cosmological parameter fits.

\begin{table}
\begin{center}
\caption{The inverse covariance matrix $\uuline{C}^{-1}$ of the
  measurements from the WiggleZ survey data of the acoustic parameter
  $A(z)$ defined by Equation \ref{eqaz}.  We have performed these
  measurements in three overlapping redshift slices $0.2 < z < 0.6$,
  $0.4 < z < 0.8$ and $0.6 < z < 1.0$ with effective redshifts $z_{\rm
    eff} = 0.44$, $0.6$ and $0.73$, respectively.  The data vector is
  $\uline{A}_{\rm obs} = (0.474, 0.442, 0.424)$, as listed in Table
  \ref{tabbaofit}.  The chi-squared statistic for any cosmological
  model vector $\uline{A}_{\rm mod}$ can be obtained via the matrix
  multiplication $\chi^2 = (\uline{A}_{\rm obs} - \uline{A}_{\rm
    mod})^T \uuline{C}^{-1} (\uline{A}_{\rm obs} - \uline{A}_{\rm
    mod})$.  The matrix is symmetric; we just quote the upper
  diagonal.}
\label{tabwigcov}
\begin{tabular}{cccc}
\hline
Redshift slice & $0.2<z<0.6$ & $0.4<z<0.8$ & $0.6<z<1.0$ \\
\hline
$0.2<z<0.6$ & $1040.3$ & $-807.5$ & $336.8$ \\
$0.4<z<0.8$ & & $3720.3$ & $-1551.9$ \\
$0.6<z<1.0$ & & & $2914.9$ \\
\hline
\end{tabular}
\end{center}
\end{table}

\subsection{Comparison to mock galaxy catalogue}

As a further test for systematic errors in our distance scale
measurements we fitted our BAO models to a dark matter halo catalogue
generated as part of the Gigaparsec WiggleZ (GiggleZ) simulation suite
(Poole et al., in prep.).  The main GiggleZ simulation consists of a
$2160^3$ particle dark matter N-body calculation in a box of side 1
$h^{-1}$ Gpc.  The cosmological parameters used for the simulation
initial conditions were $[\Omega_{\rm m}, \Omega_{\rm b}, n_{\rm s},
  h, \sigma_8] = [0.273, 0.0456, 0.96, 0.705, 0.812]$.

We measured the redshift-space correlation function of a mass-limited
subset of the dark matter halo catalogue extracted from the $z=0.6$
snapshot.  This subset of dark matter haloes, spanning a small range
of maximum circular velocities around 125 km/s, was selected to
possess a similar large-scale clustering amplitude to the WiggleZ
galaxies at that redshift.  We obtained the covariance matrix of the
measurement using jack-knife techniques.  We fitted our default
correlation function model described in Section \ref{secmod} to the
result, varying $\Omega_{\rm m} h^2$, $\alpha$, $\sigma_v$ and $b^2$
and using the same fitting range as the WiggleZ measurement, $10 < s <
180 \, h^{-1}$ Mpc.

\begin{figure}
\begin{center}
\resizebox{8cm}{!}{\rotatebox{270}{\includegraphics{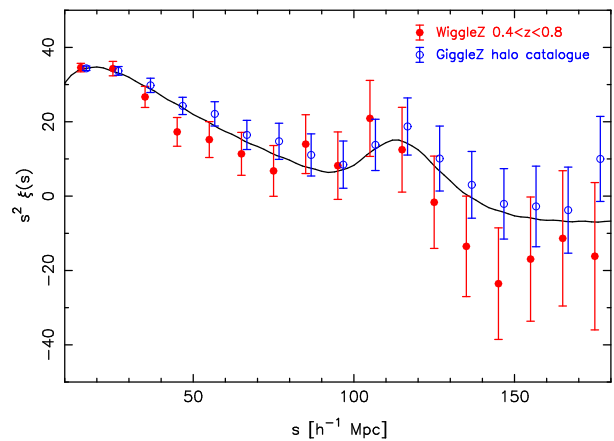}}}
\end{center}
\caption{Measurement of the galaxy correlation function $\xi(s)$ from
  a GiggleZ redshift-space halo subset at $z=0.6$, chosen to possess a
  similar large-scale clustering amplitude to the WiggleZ galaxies at
  that redshift.  We plot the correlation function in the combination
  $s^2 \, \xi(s)$ where $s$ is the co-moving redshift-space
  separation, and compare the result to the WiggleZ correlation
  function for the redshift range $0.4 < z < 0.8$.  The best-fitting
  clustering model to the GiggleZ measurement, varying the parameters
  $\Omega_{\rm m} h^2$, $\alpha$, $\sigma_v$ and $b^2$ as described in
  Section \ref{secmod}, is overplotted as the solid line.}
\label{figxigigglez}
\end{figure}

\begin{figure}
\begin{center}
\resizebox{8cm}{!}{\rotatebox{270}{\includegraphics{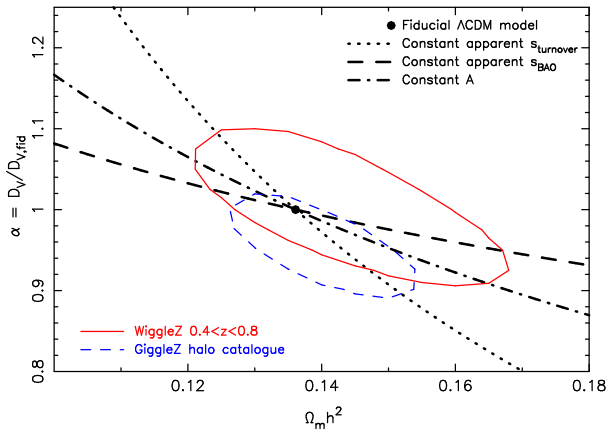}}}
\end{center}
\caption{Probability contours of the physical matter density
  $\Omega_{\rm m} h^2$ and scale distortion parameter $\alpha$
  obtained by fitting the default correlation function model to the
  GiggleZ halo subset at $z=0.6$.  We compare the result to the
  WiggleZ measurement in the redshift range $0.4 < z < 0.8$ and
  overplot the same degeneracy directions as shown in Figure
  \ref{figprobwigglez}.  The solid circle represents the location of
  our fiducial cosmological model.  The contour level in each case
  encloses regions containing $68.27\%$ of the total likelihood.}
\label{figprobgigglez}
\end{figure}

Figure \ref{figxigigglez} shows the $z=0.6$ GiggleZ halo correlation
function measurement compared to the WiggleZ correlation function for
the redshift range $0.4 < z < 0.8$ (which was plotted in the top
right-hand panel of Figure \ref{figxiwigglez}).  We overplot the
best-fitting default correlation function model for the GiggleZ data.
The 2D probability contours for $\Omega_{\rm m} h^2$ and $\alpha$ are
displayed in Figure \ref{figprobgigglez}, again compared to the $0.4 <
z < 0.8$ WiggleZ measurement and indicating the same degeneracy
directions as shown in Figure \ref{figprobwigglez}.  We conclude that
the best-fitting parameter values are consistent with the input values
of the simulation (within the statistical error expected in a
measurement that uses a single realization) and there is no evidence
for significant systematic error.  We note that the effective volume
of the halo catalogue is slightly greater than that of the WiggleZ
survey redshift range $0.4 < z < 0.8$, hence the BAO measurements are
more accurate in the case of GiggleZ.

\section{Baryon acoustic peak measurement from the full Sloan Digital Sky Survey Luminous Red Galaxy sample}
\label{seclrg}

In this Section we measure and fit the correlation function of the
SDSS-LRG DR7-Full sample.  This analysis is similar to that performed
by Kazin et al.\ (2010a) for quasi-volume-limited sub-samples with $z
< 0.36$, but now extended to a higher maximum redshift $z = 0.44$.  We
note that we assume a fiducial cosmology $\Omega_{\rm m} = 0.25$ for
this analysis, motivated by the cosmological parameters used in the
LasDamas simulations (which we use to determine the covariance matrix
of the measurement as described below in Section \ref{seclasdamas}).
The choice instead of $\Omega_{\rm m} = 0.27$, as used for the 6dFGS
and WiggleZ analyses, would yield very similar results because the
Alcock-Paczynski distortion between these cases is negligible compared
to the statistical errors in $\alpha$.

\subsection{Correlation function measurement}

\begin{figure*}
\includegraphics[width=0.5\textwidth]{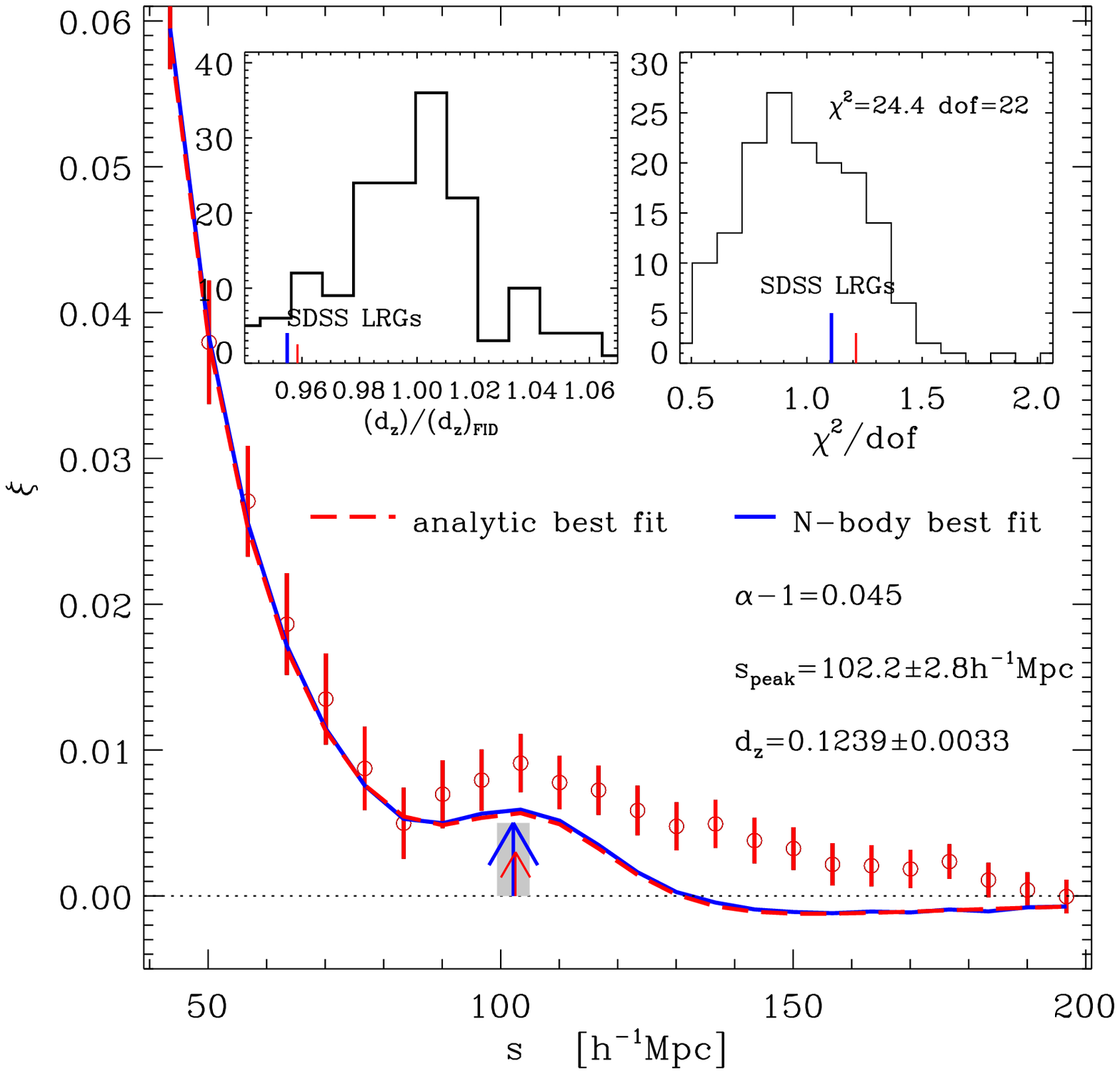}\includegraphics[width=0.5\textwidth]{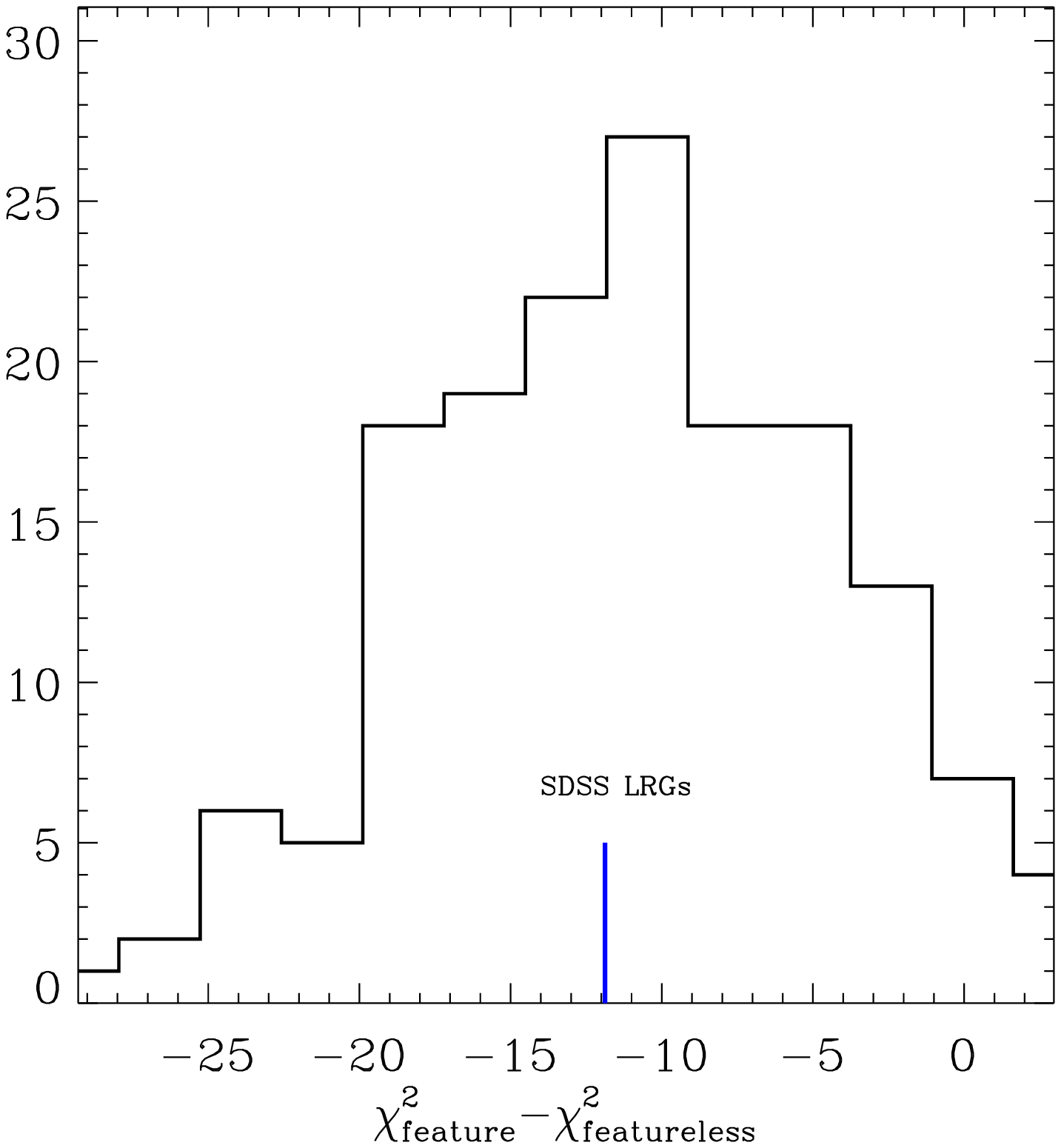}
\caption{The left-hand plot displays our correlation function
  measurement for the SDSS-LRG DR7-Full sample over the separation
  range $40 < s < 200 \, h^{-1}$ Mpc (where the error bars are the
  diagonal elements of the covariance matrix determined from 160 mock
  realizations). The solid line is the best-fitting model based on the
  mock-mean correlation function $\xi_{\rm mean}$, and the dashed line
  is the best-fitting analytic perturbation-theory model correlation
  function $\xi_{\rm pt}$ based on Equation \ref{eqxipt}.  The arrows
  point to the most likely peak position according to each model,
  where the longer arrow corresponds to the $\xi_{\rm mean}$ result
  $s_{\rm peak} = 102.2 \pm 2.8 \, h^{-1}$ Mpc.  In the right-hand
  inset the reduced chi-squared statistic $\chi^2/$dof $= 1.1 \,
  (1.2)$ using $\xi_{\rm mean}$ ($\xi_{\rm pt}$) for $22$ degrees of
  freedom is compared with a histogram of the results fitting to the
  160 individual mock realizations.  The left-hand inset compares the
  measurement of $d_{z=0.314}$ to the distribution found from the
  mocks; the offset of the measured result is due to the fact that the
  fiducial matter density $\Omega_{\rm m} = 0.25$ used to generate the
  mocks is a little lower than the current best fits to cosmological
  data.  The right-hand plot shows the distribution amongst the 160
  mocks of the difference in the chi-squared statistic between a model
  containing the baryon acoustic peak and a featureless model.  The
  $3.4$-$\sigma$ detection of the baryon acoustic feature that we find
  in DR7-Full ($\Delta \chi^2 = 11.9$) falls well within the
  distribution of values found by applying a similar analysis to the
  mock catalogues.}
\label{figxilrg}
\end{figure*}

We measured the correlation function of the SDSS-LRG DR7-Full sample
by applying the estimator of Equation \ref{eqxiest}, using random
catalogues constructed in the manner described in detail by Kazin et
al.\ (2010a).  For the purposes of the model fits in this Section we
used separation bins of width $6.6 \, h^{-1}$ Mpc spanning the range
$40 < s < 200 \, h^{-1}$ Mpc, although we also determined results in
$10 \, h^{-1}$ Mpc bins in order to combine with the 6dFGS and WiggleZ
correlation functions in Section \ref{secstack} below.  The
measurement of the DR7-Full correlation function in $6.6 \, h^{-1}$
Mpc bins is displayed in the left-hand panel of Figure \ref{figxilrg},
where the error bars are determined from the diagonal elements of the
covariance matrix of 160 mock realizations, generated as described
below in Section \ref{seclasdamas}.  The solid and dashed lines in
Figure \ref{figxilrg} are two best-fitting models, determined as
explained below in Section \ref{seclrgximod}.

The correlation function measurements in the separation range $120 < s
< 190 \, h^{-1}$ Mpc are higher than expected in the best-fitting
model.  However, it is important to remember that these data points
are correlated.  The reduced chi-squared statistics corresponding to
these models are $\chi^2$/dof $= 1.1 - 1.2$ (for 22 degrees of
freedom), which fall well within the distribution of $\chi^2$ found in
individual fits to the 160 mock catalogues, as shown in the right-hand
inset in Figure \ref{figxilrg}.  Kazin et al.\ (2010a) discussed the
excess clustering measurement in SDSS-LRG subsamples and suggested
that this is likely to result from sample variance.  This is now
reinforced by the fact that the independent-volume measurements from
the WiggleZ and 6dFGS samples do not show similar trends of excess
(see Figure \ref{figxiall}).

A potential cause of the stronger-than-expected clustering of LRGs on
large scales is the effect of not masking faint stars on
random-catalogue generation.  Ross et al.\ (2011) showed that apparent
excess large-scale angular clustering measured in photometric LRG
samples (Blake et al.\ 2007, Padmanabhan et al.\ 2007, Thomas et
al.\ 2011) is a systematic effect imprinted by anti-correlations
between faint stars and the galaxies, that can be corrected for by
masking out regions around the stars.  However, in the sparser
SDSS-DR7 LRG sample the faint stars are uncorrelated with the galaxies
at the angles of interest and do not introduce significant systematic
errors in the measured correlation function (A.S{\'a}nchez, private
communication).

\subsection{LasDamas mock galaxy catalogues}
\label{seclasdamas}

We simulated the SDSS-LRG correlation function measurement and
determined its covariance matrix using the mock galaxy catalogues
provided by the Large Suite of Dark Matter Simulations (LasDamas,
McBride et al.\ in prep.).  These N-body simulations were generated
using cosmological parameters consistent with the WMAP 5-year fits to
the CMB fluctuations (Komatsu et al.\ 2009): $[\Omega_{\rm m},
  \Omega_{\rm b}, n_{\rm s}, h, \sigma_8] = [0.25, 0.04, 1.0, 0.7,
  0.8]$.

The LasDamas collaboration generated realistic LRG mock
catalogues\footnote{http://lss.phy.vanderbilt.edu/lasdamas/} by
placing galaxies inside dark matter halos using a Halo Occupation
Distribution (HOD; Berlind \& Weinberg 2002).  The HOD parameters were
chosen to reproduce the observed galaxy number density as well as the
projected two-point correlation function $w_{\rm p}(r_{\rm p})$ of the
SDSS-LRG sample at separations $0.3 < r_{\rm p} < 30 \, h^{-1}$ Mpc.
We used a suite of 160 LRG mock catalogues constructed from light cone
samples with a mean number density $\bar{n} \sim 10^{-4} \, h^3$
Mpc$^{-3}$.  Each DR7-Full mock catalogue covers the redshift range
$0.16 < z < 0.44$ and reproduces the SDSS angular mask, corresponding
to a total volume $1.2 \, h^{-3}$ Gpc$^3$.  The mock catalogues were
subsampled to match the observed redshift distribution of the LRGs.

\subsection{Correlation function modelling}
\label{seclrgximod}

We extracted the scale of the baryon acoustic feature in the DR7-Full
correlation function measurement by fitting for the scale distortion
parameter $\alpha$ relative to a template correlation function
$\xi_{\rm fid}$ using Equation \ref{eqximod}, fitting over the
separation range $40 < s < 200 \, h^{-1}$ Mpc.  Together with the two
correlation function models already described in Section \ref{secmod},
the availability of the suite of LasDamas mock catalogues allows us to
add a third template to use as $\xi_{\rm fid}$: the mock-mean
correlation function $\xi_{\rm mean}$ of all $160$ realizations, which
includes effects due to the non-linear growth of structure,
redshift-space distortions, galaxy bias, light-coning and the observed
3D mask.

The best-fitting model taking $\xi_{\rm fid} = \xi_{\rm mean}$,
marginalizing over the correlation function amplitude, is displayed as
the solid line in Figure \ref{figxilrg}, corresponding to $\alpha =
1.045$.  The $\chi^2$ statistic of the best fit is $24.2$ (for 22
degrees of freedom).  The most likely baryon acoustic peak position
(determined using the method of Kazin et al.\ 2010a) is $s_{\rm peak}
= 102.2 \pm 2.8 \, h^{-1}$ Mpc (represented by the large arrow in
Figure \ref{figxilrg}), where the quoted error is based on the sample
variance determined by performing the same analysis on all 160 mock
catalogues.  The corresponding measurement of the distilled BAO
parameter is $d_{z=0.314} = 0.1239 \pm 0.0033$.  The distribution of
measurements of $d_z$ for the 160 mocks is shown as the left-hand
inset in Figure \ref{figxilrg}.  We do not expect the SDSS result
(vertical lines) to coincide with unity, because of the difference
between the true and fiducial cosmological parameters.

As a comparison, we also fitted to these data the two correlation
function models described in Section \ref{secmod}, parameterized by
$(d_z, \Omega_{\rm m} h^2, \sigma_v, b^2)$.  The marginalized
measurements of $d_z$ for the two models were $0.1265 \pm 0.0048$ and
$0.1272 \pm 0.0050$, consistent with our determination based on the
mock-mean correlation function (which effectively uses fixed values of
$\Omega_{\rm m} h^2$ and $\sigma_v$).

Our best-fitting analytic perturbation-theory model due to Crocce \&
Scoccimarro (2008) is displayed as the red dashed line in the
left-hand panel of Figure \ref{figxilrg}.  In this model we find that
the best-fitting value of $s_{\rm peak}$ is correlated with
$\sigma_v$, although such changes produce offsets smaller than the
1-$\sigma$ statistical error in $\alpha$ (represented by the grey
region around the short arrows in Figure \ref{figxilrg}).

\begin{figure*}
\begin{center}
\resizebox{13cm}{!}{\rotatebox{270}{\includegraphics{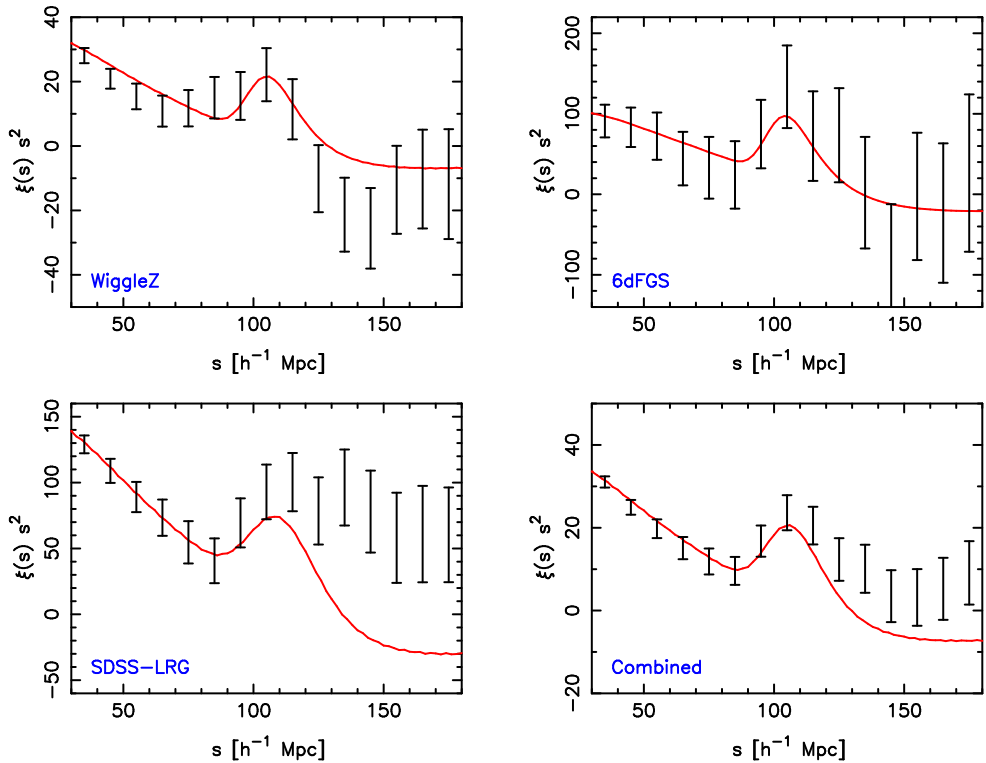}}}
\end{center}
\caption{The correlation function measurements $\xi(s)$ for the
  WiggleZ, SDSS-LRG and 6dFGS galaxy samples, plotted in the
  combination $s^2 \, \xi(s)$ where $s$ is the co-moving
  redshift-space separation. The lower right-hand panel shows the
  combination of these measuremements with inverse-variance weighting.
  The best-fitting clustering models in each case, varying the
  parameters $\Omega_{\rm m} h^2$, $\alpha$, $\sigma_v$ and $b^2$ as
  described in Section \ref{secmod}, are overplotted as the solid
  lines.}
\label{figxiall}
\end{figure*}

\subsection{Significance of detection of the SDSS-LRG baryon acoustic feature}

We assessed the statistical significance of the detection of the
baryon acoustic peak in the SDSS-LRG sample in a similar style to the
WiggleZ analysis described in Section \ref{secwigfit}, by comparing
the best-fitting values of $\chi^2$ for models containing a baryon
acoustic feature ($\chi^2_{\rm feature}$) and featureless models
($\chi^2_{\rm featureless}$) constructed using the ``no-wiggles''
power spectrum of Eisenstein \& Hu (1998).  We used the
perturbation-theory model for the baryon acoustic peak described in
Section \ref{secmod} when constructing these models.

The SDSS-LRG dataset produced $\Delta \chi^2 = \chi^2_{\rm feature} -
\chi^2_{\rm featureless}= -11.9$ over the separation range $40 < s <
200 \, h^{-1}$ Mpc, corresponding to a detection of the baryon
acoustic feature with significance of $3.4$-$\sigma$.  The histogram
resulting from repeating this analysis for all $160$ mocks is
displayed in the right-hand panel of Figure \ref{figxilrg}, following
Cabre \& Gaztanaga (2011); we see that the SDSS result is as expected
from an average realization.

We used the same method to compare the significance of detection of
the acoustic peak in DR7-Full with that obtained in the volume-limited
LRG sub-samples analyzed by Kazin et al.\ (2010a).  The sample
``DR7-Sub'', a quasi-volume-limited LRG catalogue spanning redshift
range $0.16 < z < 0.36$ and luminosity range $-23.2 < M_g < -21.2$,
yields a detection significance of $2.2$-$\sigma$.  For the sample
``DR7-Bright'', a sparser volume-limited catalogue with a brighter
luminosity cut $-23.2 < M_g < -21.8$, the significance of the baryon
acoustic feature is just below 2-$\sigma$.

\section{The stacked baryon acoustic peak}
\label{secstack}

\begin{figure*}
\begin{center}
\resizebox{13cm}{!}{\rotatebox{270}{\includegraphics{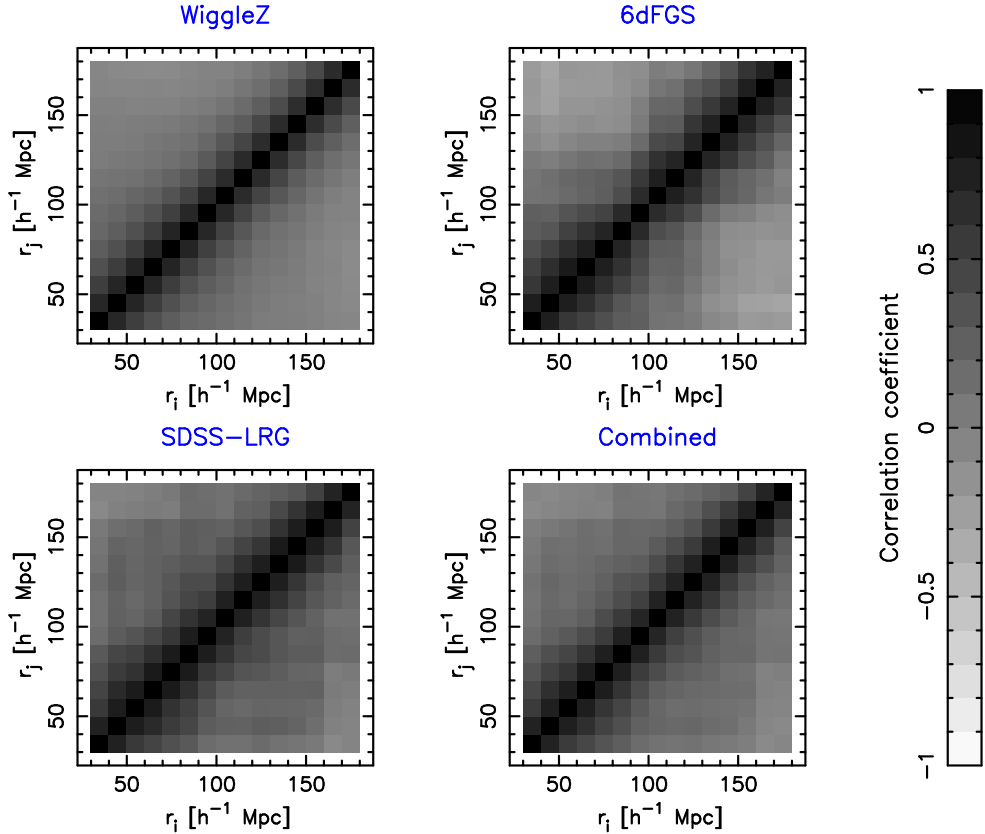}}}
\end{center}
\caption{The amplitude of the cross-correlation $C_{ij}/\sqrt{C_{ii}
    C_{jj}}$ of the covariance matrix $C_{ij}$ for the WiggleZ,
  SDSS-LRG and 6dFGS correlation functions.  The lower right-hand
  panel shows the covariance matrix of the combined correlation
  function.  The covariance matrices for the WiggleZ and 6dFGS samples
  are determined using lognormal realizations, and that of the
  SDSS-LRG sample is obtained from an ensemble of N-body simulations.
  The plot of the WiggleZ cross-correlation matrix in the upper-left
  hand panel is reproduced from the lower right-hand panel of Figure
  \ref{figcovwigglez}.}
\label{figcovall}
\end{figure*}

Our goal in this Section is to assess the overall statistical
significance with which the baryon acoustic peak is detected in the
combination of current galaxy surveys.  In order to do this we
combined the galaxy correlation functions measured from the WiggleZ
Survey, the Sloan Digital Sky Survey Luminous Red Galaxy (SDSS-LRG)
sample and the 6-degree Field Galaxy Survey (6dFGS), and fitted the
models described in Section \ref{secmod} to the result.  Although we
acknowledge that model fits to a combination of correlation functions
obtained using different redshifts and galaxy types will produce
parameter values that evade an easy physical interpretation, the
resulting statistical significance of the BAO detection remains a
quantity of interest.

\subsection{The 6dFGS baryon acoustic peak measurement}
\label{sec6df}

For completeness we summarize here the measurement of the baryon
acoustic peak from the 6dFGS reported by Beutler et al.\ (2011).
After optimal weighting of the data to minimize the correlation
function error at the baryon acoustic peak, the 6dFGS sample covered
an effective volume $V_{\rm eff} = 0.08 \, h^{-3}$ Gpc$^3$ with
effective redshift $z_{\rm eff} = 0.106$.  Beutler et al.\ fitted the
model defined by our Equation \ref{eqxipt} to the 6dFGS correlation
function, using lognormal realizations to determine the data
covariance matrix and varying the parameter set $\Omega_{\rm m} h^2$,
$\alpha$, $\sigma_v$ and $b^2$.  The model fits were performed over
the separation range $10 < s < 190 \, h^{-1}$ Mpc, with checks made
that the best-fitting parameters were not sensitive to the minimum
separation employed.  The resulting measurements of the distance scale
were quantified as $D_V(0.106) = 457 \pm 27$ Mpc, $d_{0.106} = 0.336
\pm 0.015$ or $A(0.106) = 0.526 \pm 0.028$.  The statistical
significance of the detection of the acoustic peak was estimated to be
$2.4$-$\sigma$, based on the difference in chi-squared $\Delta \chi^2
= 5.6$ between the best-fitting model and the corresponding best fit
of a zero-baryon model.

\subsection{The combined correlation function}
\label{secxicomb}

\begin{figure}
\begin{center}
\resizebox{8cm}{!}{\rotatebox{270}{\includegraphics{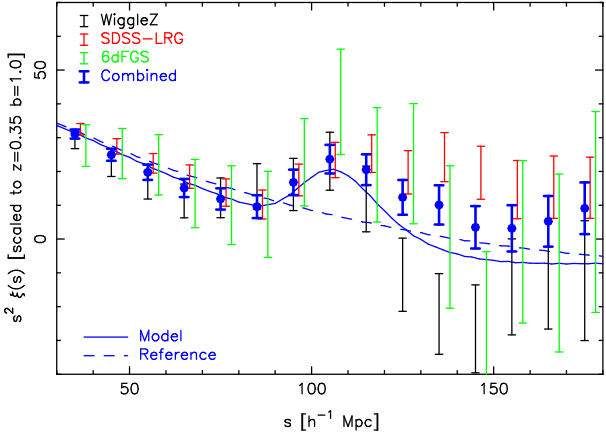}}}
\end{center}
\caption{An overplot of the correlation function measurements $\xi(s)$
  for the WiggleZ, SDSS-LRG and 6dFGS galaxy samples, plotted in the
  combination $s^2 \, \xi(s)$ where $s$ is the co-moving
  redshift-space separation.  A normalization correction has been
  applied to these correlation functions to allow for the differing
  effective redshifts and galaxy bias factors of the samples (see text
  for details).  The combined correlation function, determined by
  inverse-variance weighting, is also plotted.  The best-fitting
  clustering model to the combined correlation function (varying
  $\Omega_{\rm m} h^2$, $\alpha$, $\sigma_v$ and $b^2$) is overplotted
  as the solid line.  We also show as the dashed line the
  corresponding ``no-wiggles'' reference model (Eisenstein \& Hu
  1998), constructed from a power spectrum with the same clustering
  amplitude but lacking baryon acoustic oscillations.}
\label{figxioverplot}
\end{figure}

Figure \ref{figxiall} displays the three survey correlation functions
combined in our study: the WiggleZ $0.2 < z < 1.0$ measurement plotted
in the lower right-hand panel of Figure \ref{figxiwigglez}, the 6dFGS
correlation function reported by Beutler et al.\ (2011), and the
SDSS-LRG DR7-Full measurement described in Section \ref{seclrg} (using
a binning of $10 \, h^{-1}$ Mpc in all cases).  These correlation
functions have quite different amplitudes owing to differences between
the growth factors at the effective redshifts $z$ of the samples and
the bias factors $b$ of the various galaxy tracers.  Before stacking
these functions we make an amplitude correction to a common redshift
$z_0=0.35$ and bias factor $b_0=1$, by multiplying each correlation
function by $[b_0^2 \, G(z_0)^2 \, B_0(\beta_0)]/[b^2 \, G(z)^2 \,
  B_0(\beta)]$ where $G(z)$ is the linear growth factor at redshift
$z$ and $B_0(\beta) = 1 + \frac{2}{3} \beta + \frac{1}{5} \beta^2$ is
the Kaiser boost factor in terms of the redshift-space distortion
parameter $\beta = \Omega_{\rm m}(z)^{6/11}/b$ (Kaiser 1987).  When
calculating these quantities we assumed that the redshifts of the
WiggleZ, SDSS-LRG and 6dFGS samples were $z = (0.6, 0.314, 0.106)$ and
the bias factors were $b = (1.1, 2.2, 1.8)$.  After making these
normalization corrections we then combined the correlation functions
and their corresponding covariance matrices using inverse-variance
weighting in the same style as Equations \ref{eqxicomb} and
\ref{eqcovcomb}.  The resulting total correlation function is plotted
in the lower right-hand panel of Figure \ref{figxiall}.  The
covariance matrices of the different survey correlation functions and
final combination are displayed in Figure \ref{figcovall}.  An
additional overplot of these correlation functions is provided in
Figure \ref{figxioverplot}.  We note that although the SDSS-LRG
correlation function measurement used the fiducial cosmology
$\Omega_{\rm m} = 0.25$, compared to the choice $\Omega_{\rm m} =
0.27$ for the WiggleZ and 6dFGS analyses, the Alcock-Paczynski
distortion between these cases is negligible compared to the
statistical errors in $\alpha$.

\subsection{Significance of the detection of the baryon acoustic peak in the combined sample}

We fitted the clustering model described in Section \ref{secmod} to
the combined correlation function over separation range $30 < s < 180
\, h^{-1}$ Mpc, varying $\Omega_{\rm m} h^2$, $\alpha$, $\sigma_v$ and
$b^2$ and using an effective redshift $z = 0.35$.  We used the more
conservative minimum fitted scale $30 \, h^{-1}$ Mpc for the analysis
of the stacked correlation function in this Section, compared to $10
\, h^{-1}$ Mpc for the fits to the WiggleZ correlation function in
Section \ref{secwigglez}, because (1) the required non-linear
corrections become more important for galaxy samples such as the 6dFGS
and SDSS LRGs, which are both more biased and at lower redshift than
the WiggleZ sample, and (2) systematic errors in the fitting become
relatively more important for this combined dataset with higher
signal-to-noise.  Although we fixed the relative bias factors of the
galaxy samples when stacking the survey correlation functions in
Section \ref{secxicomb}, we still marginalized over an absolute
normalization $b^2 \sim 1$ when fitting the model in this Section.

We obtained a good fit to the stacked correlation function with
$\chi^2 = 11.3$ (for 11 degrees of freedom) and marginalized parameter
values $\Omega_{\rm m} h^2 = 0.132 \pm 0.014$, $\alpha = 1.037 \pm
0.036$ and $\sigma_v = 4.5 \pm 1.8 \, h^{-1}$ Mpc.  Although the
best-fitting value of $\alpha$ must be interpreted as some effective
value integrating over redshift, we can conclude that the measured BAO
distance scale is consistent with the fiducial model.

We quantified the significance of the detection of the acoustic peak
in the combined sample using two methods.  Firstly, we repeated the
parameter fit replacing the model correlation function with one
generated using a ``no-wiggles'' reference power spectrum (Eisenstein
\& Hu 1998).  The minimum value obtained for the $\chi^2$ statistic in
this case was $32.7$, indicating that the model containing baryon
oscillations was favoured by $\Delta \chi^2 = 21.4$.  This corresponds
to a detection of the acoustic peak with a statistical significance of
$4.6$-$\sigma$.

As an alternative approach for assessing the significance of the
detection, we changed the fiducial baryon density to $\Omega_{\rm b} =
0$ and repeated the parameter fit.  For zero baryon density we
generated the model matter power spectrum using the fitting formulae
of Eisenstein \& Hu (1998), rather than using the CAMB software.  The
minimum value obtained for the $\chi^2$ statistic was now $35.3$, this
time suggesting that the acoustic peak had been detected with a
significance of $4.9$-$\sigma$.  The reason that the significance of
detection varies between these two methods of assessment is that in
the latter case, where the baryon density is changed, the overall
shape of the clustering pattern is also providing information used to
disfavour the $\Omega_{\rm b} = 0$ model, whereas in the former case
only the presence of the acoustic peak varies between the two sets of
models.

\section{Cosmological parameter fits}
\label{seccosmofit}

In this Section we fit cosmological models to the latest distance
datasets comprising BAO, supernovae and CMB measurements.  Our aim is
to compare parameter fits to BAO+CMB data (excluding supernovae) and
SNe+CMB data (excluding BAO) as a robust check for systematic errors
in these distance probes.

\subsection{BAO dataset}

The latest BAO distance dataset, including the 6dFGS, SDSS and WiggleZ
surveys, now comprises BAO measurements at six different redshifts.
These data are summarized in Table \ref{tabbaodist}.  Firstly, we use
the measurement of $d_{0.106} = 0.336 \pm 0.015$ from the 6dFGS
reported by Beutler et al.\ (2011).  Secondly, we add the two
correlated measurements of $d_{0.2}$ and $d_{0.35}$ determined by
Percival et al.\ (2010) from fits to the power spectra of LRGs and
main-sample galaxies in the SDSS (spanning a range of wavenumbers
$0.02 < k < 0.3 \, h$ Mpc$^{-1}$).  The correlation coefficient for
these last two measurements is $0.337$.  We note that our own LRG
baryon acoustic peak measurements reported above in Section
\ref{seclrg} are entirely consistent with these fits.  Finally, we
include the three correlated measurements of $A(z=0.44)$, $A(z=0.6)$
and $A(z=0.73)$ reported in this study, using the inverse covariance
matrix listed in Table \ref{tabwigcov}.

In our cosmological model fitting we assume that the BAO distance
errors are Gaussian in nature.  Modelling potential non-Gaussian tails
in the likelihood is beyond the scope of this paper, although we note
that they may not be negligible (Percival et al.\ 2007, Percival et
al.\ 2010, Bassett \& Afshordi 2010).  We caution that the 2-$\sigma$
confidence regions displayed in the Figures in this Section might not
necessarily follow the Gaussian scaling.  The WiggleZ and SDSS-LRG
surveys share a sky overlap of $\approx 500$ deg$^2$ for redshift
range $z < 0.5$; given that the SDSS-LRG measurement is derived across
a sky area $\approx 8000$ deg$^2$ and the errors in both measurements
contain a significant component due to shot noise, the resulting
covariance is negligible.

\begin{figure*}
\begin{center}
\resizebox{13cm}{!}{\rotatebox{270}{\includegraphics{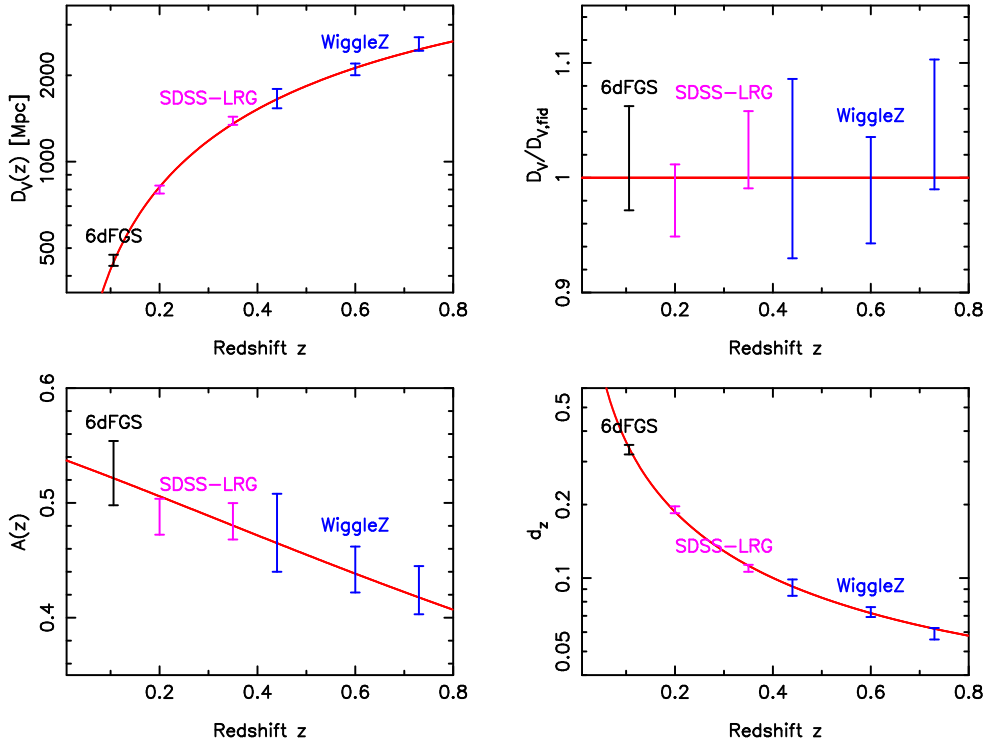}}}
\end{center}
\caption{Current measurements of the cosmic distance scale using the
  BAO standard ruler applied to the 6dFGS, SDSS and WiggleZ surveys
  (where the data is taken from Beutler et al.\ 2011, Percival et
  al.\ 2010 and this study).  The results are compared to a flat
  $\Lambda$CDM cosmological model with matter density $\Omega_{\rm m}
  = 0.29$ and Hubble parameter $h = 0.69$.  Various representations of
  the data are shown: the BAO distance $D_V(z)$ recovered from fits to
  the angle-averaged clustering measurements (top left-hand panel),
  these distances ratioed to the fiducial model (top right-hand
  panel), the distilled parameter $A(z)$ (defined by Equation
  \ref{eqaz}) extracted from fits governed by both the acoustic peak
  and clustering shape (bottom left-hand panel), and the distilled
  parameter $d_z$ determined by fits controlled by solely the acoustic
  peak information (bottom right-hand panel).  We note that the
  conversion of the BAO fits to the measurements of $D_V(z)$ presented
  in the upper two plots requires a value for the standard ruler scale
  to be assumed: we take $r_s(z_d) = 152.40$ Mpc, obtained using
  Equation 6 in Eisenstein \& Hu (1998) evaluated for our fiducial
  model $\Omega_{\rm m} h^2 = 0.1381$ and $\Omega_{\rm b} h^2 =
  0.02227$.}
\label{figbaodist}
\end{figure*}

This BAO distance dataset is plotted in Figure \ref{figbaodist}
relative to a flat $\Lambda$CDM cosmological model with matter density
$\Omega_{\rm m} = 0.29$ and Hubble parameter $h = 0.69$ (these values
provide the best fit to the combined cosmological datasets as
discussed below).  The panels of Figure \ref{figbaodist} show various
representations of the BAO dataset including $D_V(z)$ and the
distilled parameters $A(z)$ and $d_z$.

\begin{table}
\begin{center}
\caption{The BAO distance dataset from the 6dFGS, SDSS and WiggleZ
  surveys.  Measurements of the distilled parameters $d_z$ and $A(z)$
  are quoted.  The most appropriate choices to be used in cosmological
  parameter fits are indicated by bold font.  For the SDSS data, the
  values of $A(z)$ are obtained by scaling from the measurements of
  $d_z$ reported by Percival et al.\ (2010) using their fiducial
  cosmological parameters and the same fractional error.  The pairs of
  measurements at $z=(0.2,0.35)$, $z=(0.44,0.6)$ and $z=(0.6,0.73)$
  are correlated with coefficients $0.337$, $0.369$ and $0.438$,
  respectively.  The inverse covariance matrix of the data points at
  $z=(0.2,0.35)$ is given by Equation 5 in Percival et al.\ (2010).
  The inverse covariance matrix of the data points at
  $z=(0.44,0.6,0.73)$ is given in Table \ref{tabwigcov} above.  The
  other measurements are uncorrelated.}
\label{tabbaodist}
\begin{tabular}{cccc}
\hline
Sample & $z$ & $d_z$ & $A(z)$ \\
\hline
6dFGS & $0.106$ & ${\mathbf 0.336 \pm 0.015}$ & $0.526 \pm 0.028$ \\
SDSS & $0.2$ & ${\mathbf 0.1905 \pm 0.0061}$ & $0.488 \pm 0.016$ \\
SDSS & $0.35$ & ${\mathbf 0.1097 \pm 0.0036}$ & $0.484 \pm 0.016$ \\
WiggleZ & $0.44$ & $0.0916 \pm 0.0071$ & ${\mathbf 0.474 \pm 0.034}$ \\
WiggleZ & $0.6$ & $0.0726 \pm 0.0034$ & ${\mathbf 0.442 \pm 0.020}$ \\
WiggleZ & $0.73$ & $0.0592 \pm 0.0032$ & ${\mathbf 0.424 \pm 0.021}$ \\
\hline
\end{tabular}
\end{center}
\end{table}

\subsection{SNe dataset}

We used the ``Union 2'' compilation by Amanullah et al.\ (2010) as our
supernova dataset, obtained from the website {\tt
  http://supernova.lbl.gov/Union}.  This compilation of 557 supernovae
includes data from Hamuy et al.\ (1996), Riess et al.\ (1999, 2007),
Astier et al.\ (2006), Jha et al.\ (2006), Wood-Vasey et al.\ (2007),
Holtzman et al.\ (2008), Hicken et al.\ (2009) and Kessler et
al.\ (2009).  The data is represented as a set of values of the
distance modulus for each supernova
\begin{equation}
\mu = 5 \, {\rm log}_{10} \left[ \frac{D_L(z)}{1 \, {\rm Mpc}} \right]
+ 25 ,
\end{equation}
where $D_L(z)$ is the luminosity distance at redshift $z$.  The values
of $\mu$ are reported for a particular choice of the normalization $M
- 5 \, {\rm log}_{10} h$, which is marginalized as an unknown
parameter in our analysis as described below.  When fitting
cosmological models to these SNe data we used the full covariance
matrix of these measurements including systematic errors, as reported
by Amanullah et al.\ (2010).

\begin{figure}
\begin{center}
\resizebox{8cm}{!}{\rotatebox{270}{\includegraphics{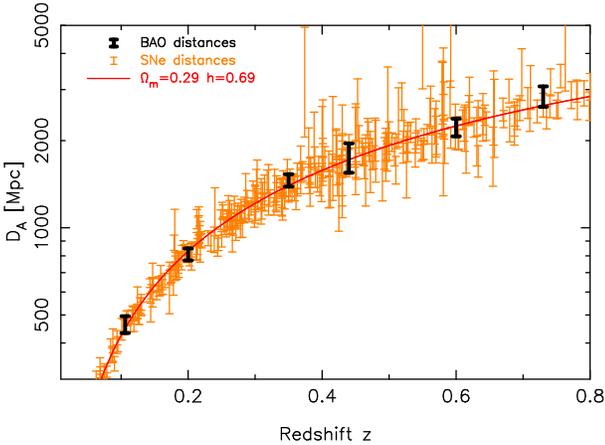}}}
\end{center}
\caption{Comparison of the accuracy with which supernovae and baryon
  acoustic oscillations map out the cosmic distance scale at $z <
  0.8$.  For the purposes of this Figure, BAO measurements of $D_V(z)$
  have been converted into $D_A(z)$ assuming a Hubble parameter $H(z)$
  for a flat $\Lambda$CDM model with $\Omega_{\rm m} = 0.29$ and $h =
  0.69$, indicated by the solid line in the Figure, and SNe
  measurements of $D_L(z)$ have been plotted assuming $D_A(z) =
  D_L(z)/(1+z)^2$.}
\label{figcomparedist}
\end{figure}

Figure \ref{figcomparedist} is a representation of the consistency and
relative accuracy with which baryon oscillation measurements and
supernovae currently map out the cosmic distance scale.  In order to
construct this figure we converted the BAO measurements of $D_V(z)$
into $D_A(z)$ assuming a Hubble parameter $H(z)$ for a flat
$\Lambda$CDM model with $\Omega_{\rm m} = 0.29$ and $h = 0.69$.  The
binned supernovae data currently measure the distance-redshift
relation at $z < 0.8$ with $3-4$ times higher accuracy than the BAOs,
although we note that the consequences for cosmological parameter fits
are highly influenced by the differing normalization of the two
methods.  The supernovae measure the relative luminosity distance to
the relation at $z=0$, $D_L(z) H_0/c$, owing to the unknown value of
the standard-candle absolute magnitude $M$.  The BAOs measure a
distance scale relative to the sound horizon at baryon drag calibrated
by the CMB data, effectively an absolute measurement of $D_V(z)$ given
that the error is dominated by the statistical uncertainty in the
clustering fits, rather than any systematic uncertainty in the sound
horizon calibration from the CMB.

When undertaking cosmological fits to the supernovae dataset, we
performed an analytic marginalization over the unknown absolute
normalization $M - 5 \, {\rm log}_{10} h$ (Goliath et al.\ 2001,
Bridle et al.\ 2002).  This is carried out by determining the
chi-squared statistic for each cosmological model as
\begin{equation}
\chi^2 = \uline{y}^T \, \uuline{C}_{\rm SN}^{-1} \, \uline{y} -
\frac{(\sum_{ij} C_{{\rm SN},ij}^{-1} \, y_j)^2}{\sum_{ij} C_{{\rm
      SN},ij}^{-1}}
\end{equation}
where $\uline{y}$ is the vector representing the difference between
the distance moduli of the data and model, and $\uuline{C}_{\rm
    SN}^{-1}$ is the inverse covariance matrix for the supernovae
distance moduli.

\subsection{CMB dataset}

We included the CMB data in our cosmological fits using the Wilkinson
Microwave Anisotropy Probe (WMAP) ``distance priors'' (Komatsu et
al.\ 2009) using the 7-year WMAP results reported by Komatsu et
al.\ (2011).  The distance priors quantify the complete CMB likelihood
via a 3-parameter covariance matrix for the acoustic index $\ell_A$,
the shift parameter ${\mathcal R}$ and the redshift of recombination
$z_*$, as given in Table 10 of Komatsu et al.\ (2011).  When deriving
these quantities we assumed a physical baryon density $\Omega_{\rm b}
h^2 = 0.02227$, a CMB temperature $T_{\rm CMB} = 2.725 K$ and a number
of relativistic degrees of freedom $N_{\rm eff} = 3.04$.

\subsection{Flat $w$ models}

We first fitted a flat $w$CDM cosmological model in which spatial
curvature is fixed at $\Omega_{\rm k} = 0$ but the equation-of-state
$w$ of dark energy is varied as a free parameter.  We fitted for the
three parameters $(\Omega_{\rm m}, \Omega_{\rm m} h^2, w)$ using flat,
wide priors which extend well beyond the regions of high likelihood
and have no effect on the cosmological fits.  The best-fitting model
has $\chi^2 = 532.9$ for 563 degrees of freedom, representing a good
fit to the distance dataset.

\begin{figure}
\begin{center}
\resizebox{8cm}{!}{\rotatebox{270}{\includegraphics{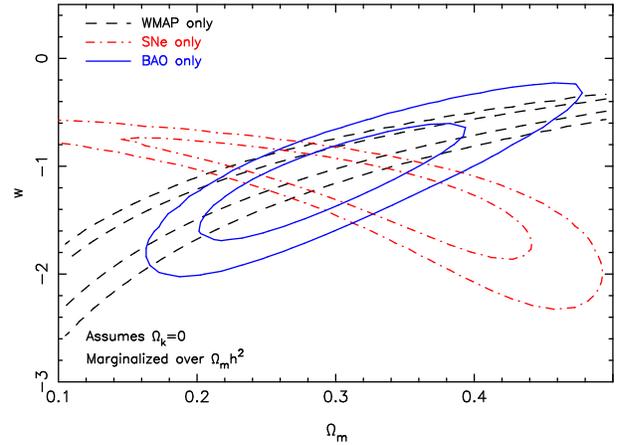}}}
\end{center}
\caption{The joint probability for parameters $\Omega_{\rm m}$ and $w$
  fitted separately to the WMAP, BAO and SNe distance data,
  marginalized over $\Omega_{\rm m} h^2$ and assuming $\Omega_{\rm k}
  = 0$.  The two contour levels in each case enclose regions
  containing $68.27\%$ and $95.45\%$ of the total likelihood.}
\label{figprobomwcomp}
\end{figure}

\begin{figure}
\begin{center}
\resizebox{8cm}{!}{\rotatebox{270}{\includegraphics{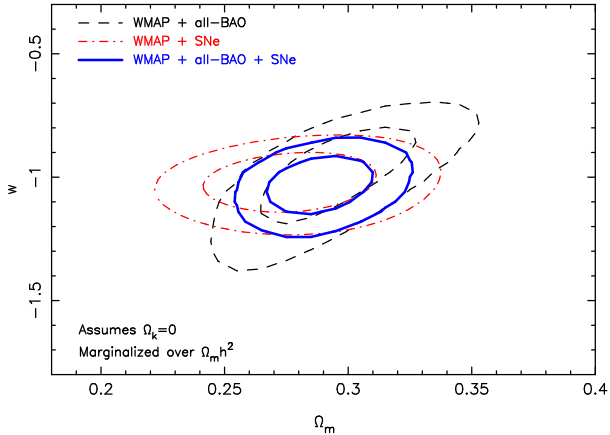}}}
\end{center}
\caption{The joint probability for parameters $\Omega_{\rm m}$ and $w$
  fitted to various combinations of WMAP, BAO and SNe distance data,
  marginalized over $\Omega_{\rm m} h^2$ and assuming $\Omega_{\rm k}
  = 0$.  The two contour levels in each case enclose regions
  containing $68.27\%$ and $95.45\%$ of the total likelihood.}
\label{figprobomwall}
\end{figure}

Figures \ref{figprobomwcomp} and \ref{figprobomwall} compare the joint
probability of $\Omega_{\rm m}$ and $w$, marginalizing over
$\Omega_{\rm m} h^2$, for the individual WMAP, BAO and SNe datasets
along with various combinations.  We note that for the ``BAO only''
contours in Figure \ref{figprobomwcomp}, we have not used any CMB
calibration of the standard ruler scale $r_s(z_d)$, and thus the 6dFGS
and SDSS measurements of $d_z = r_s(z_d)/D_V(z)$ do not contribute
strongly to these constraints.  Hence the addition of the CMB data in
Figure \ref{figprobomwall} has the benefit of both improving the
information from the $d_z$ measurements by determining $r_s(z_d)$, and
contributing the WMAP distance prior constraints.  The WMAP+BAO and
WMAP+SNe data produce consistent determinations of the cosmological
parameters, with the error in the equation-of-state $\Delta w \approx
0.1$.  Combining all three datasets produces the marginalized result
$w = -1.034 \pm 0.080$ (errors in the other parameters are listed in
Table \ref{tabcosmofit}; the quoted error in $h$ results from fitting
the three parameters $\Omega_{\rm m}$, $h$ and $w$).  The best-fitting
equation-of-state is consistent with a cosmological constant model for
which $w = -1$.

We caution that the probability contours plotted in Figures
\ref{figprobomwcomp} and \ref{figprobomwall} (and other similar
Figures in this Section) assume that the errors in the BAO distance
dataset are Gaussian.  If the likelihood contains a significant
non-Gaussian tail, the 2-$\sigma$ region could be affected.

We repeated the WMAP+BAO fit comparing the two different
implementations of the SDSS-LRG BAO distance-scale measurements: the
Percival et al.\ (2010) power spectrum fitting at $z=0.2$ and
$z=0.35$, and our correlation function fit presented in Section
\ref{seclrg}.  We found that the marginalized measurements of $w$ in
the two cases were $-1.00 \pm 0.13$ and $-0.97 \pm 0.13$,
respectively.  Our results are therefore not significantly changed by
the methodology used for these LRG fits.

\subsection{Curved $\Lambda$ models}

We next fitted a curved $\Lambda$CDM model, in which we fixed the
equation-of-state of dark energy at $w = -1$ but added the spatial
curvature $\Omega_{\rm k}$ as an additional free parameter.  We fitted
for the three parameters $(\Omega_{\rm m}, \Omega_{\rm m} h^2,
\Omega_{\rm k})$ using flat, wide priors which extend well beyond the
regions of high likelihood and have no effect on the cosmological
fits.  The best-fitting model has $\chi^2 = 532.7$ for 563 degrees of
freedom.

\begin{figure}
\begin{center}
\resizebox{8cm}{!}{\rotatebox{270}{\includegraphics{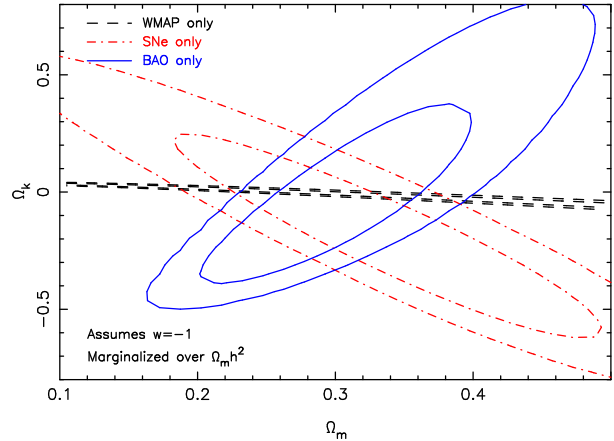}}}
\end{center}
\caption{The joint probability for parameters $\Omega_{\rm m}$ and
  $\Omega_{\rm k}$ fitted separately to the WMAP, BAO and SNe distance
  data, marginalized over $\Omega_{\rm m} h^2$ and assuming $w = -1$.
  The two contour levels in each case enclose regions containing
  $68.27\%$ and $95.45\%$ of the total likelihood.}
\label{figprobomokcomp}
\end{figure}

\begin{figure}
\begin{center}
\resizebox{8cm}{!}{\rotatebox{270}{\includegraphics{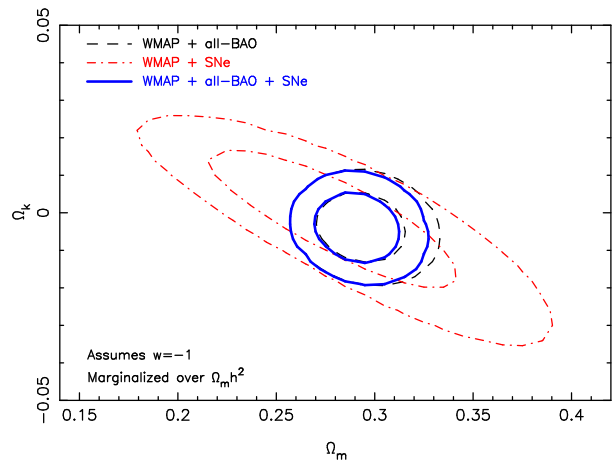}}}
\end{center}
\caption{The joint probability for parameters $\Omega_{\rm m}$ and
  $\Omega_{\rm k}$ fitted to various combinations of WMAP, BAO and SNe
  distance data, marginalized over $\Omega_{\rm m} h^2$ and assuming
  $w = -1$.  The two contour levels in each case enclose regions
  containing $68.27\%$ and $95.45\%$ of the total likelihood.}
\label{figprobomokall}
\end{figure}

Figures \ref{figprobomokcomp} and \ref{figprobomokall} compare the
joint probability of $\Omega_{\rm m}$ and $\Omega_{\rm k}$,
marginalizing over $\Omega_{\rm m} h^2$, for the individual WMAP, BAO
and SNe datasets along with various combinations.  Once more, we find
that fits to WMAP+BAO and WMAP+SNe produce mutually consistent
results.  The BAO data has higher sensitivity to curvature because of
the long lever arm represented by the relation of distance
measurements at $z < 1$ and at recombination.  Combining all three
datasets produces the marginalized result $\Omega_{\rm k} = -0.0040
\pm 0.0062$ (errors in the other parameters are listed in Table
\ref{tabcosmofit}).  The best-fitting parameters are consistent with
zero spatial curvature.

\subsection{Additional degrees of freedom}

\begin{figure}
\begin{center}
\resizebox{8cm}{!}{\rotatebox{270}{\includegraphics{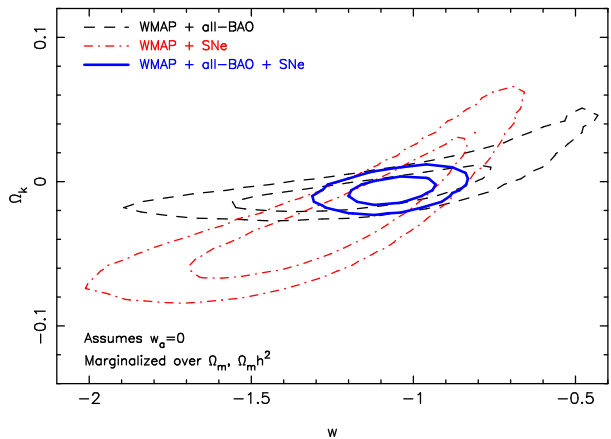}}}
\end{center}
\caption{The joint probability for parameters $\Omega_{\rm k}$ and $w$
  fitted to various combinations of WMAP, BAO and SNe distance data,
  marginalized over $\Omega_{\rm m}$ and $\Omega_{\rm m} h^2$.  The
  two contour levels in each case enclose regions containing $68.27\%$
  and $95.45\%$ of the total likelihood.}
\label{figprobwok}
\end{figure}

We fitted two further cosmological models, each containing an
additional parameter.  Firstly we fitted a curved $w$CDM model in
which we varied both the dark energy equation-of-state and the spatial
curvature as free parameters.  The best-fitting model has $\chi^2 =
531.9$ for 562 degrees of freedom, representing an improvement of
$\Delta \chi^2 = 1.0$ compared to the case where $\Omega_{\rm k} = 0$,
for the addition of a single extra parameter.  In terms of information
criteria this does not represent a sufficient improvement to justify
the addition of the extra degree of freedom.  Figure \ref{figprobwok}
compares the joint probability of $w$ and $\Omega_{\rm k}$,
marginalizing over $\Omega_{\rm m}$ and $\Omega_{\rm m} h^2$, for the
three cases WMAP+BAO, WMAP+SNe and WMAP+BAO+SNe.  Combining all three
datasets produces the marginalized measurements $w = -1.063 \pm 0.094$
and $\Omega_{\rm k} = -0.0061 \pm 0.0070$.

We finally fitted a flat $w(a)$CDM cosmological model in which spatial
curvature is fixed at $\Omega_{\rm k} = 0$ but the equation-of-state
of dark energy is allowed to vary with scale factor $a$ in accordance
with the Chevallier-Polarski-Linder parameterization $w(a) = w_0 +
(1-a)w_a$ (Chevallier \& Polarski 2001, Linder 2003).  The
best-fitting model has $\chi^2 = 531.9$ for 562 degrees of freedom,
and again the improvement in the value of $\chi^2$ compared to the
case where $w_a = 0$ does not justify the addition of the extra degree
of freedom.  Combining all three datasets produces the marginalized
measurements $w_0 = -1.09 \pm 0.17$ and $w_a = 0.19 \pm 0.69$.  We
note that the addition of the BAO measurements to the WMAP+SNe dataset
produces a more significant improvement for fits involving
$\Omega_{\rm k}$ than for $w_a$.

In all cases, the best-fitting parameters are consistent with a flat
cosmological constant model for which $w_0 = -1$, $w_a = 0$ and
$\Omega_{\rm k} = 0$.  The best-fitting values and errors in the
parameters for the various models, for the fits using all three
datasets, are listed in Table \ref{tabcosmofit}.

\begin{figure}
\begin{center}
\resizebox{8cm}{!}{\rotatebox{270}{\includegraphics{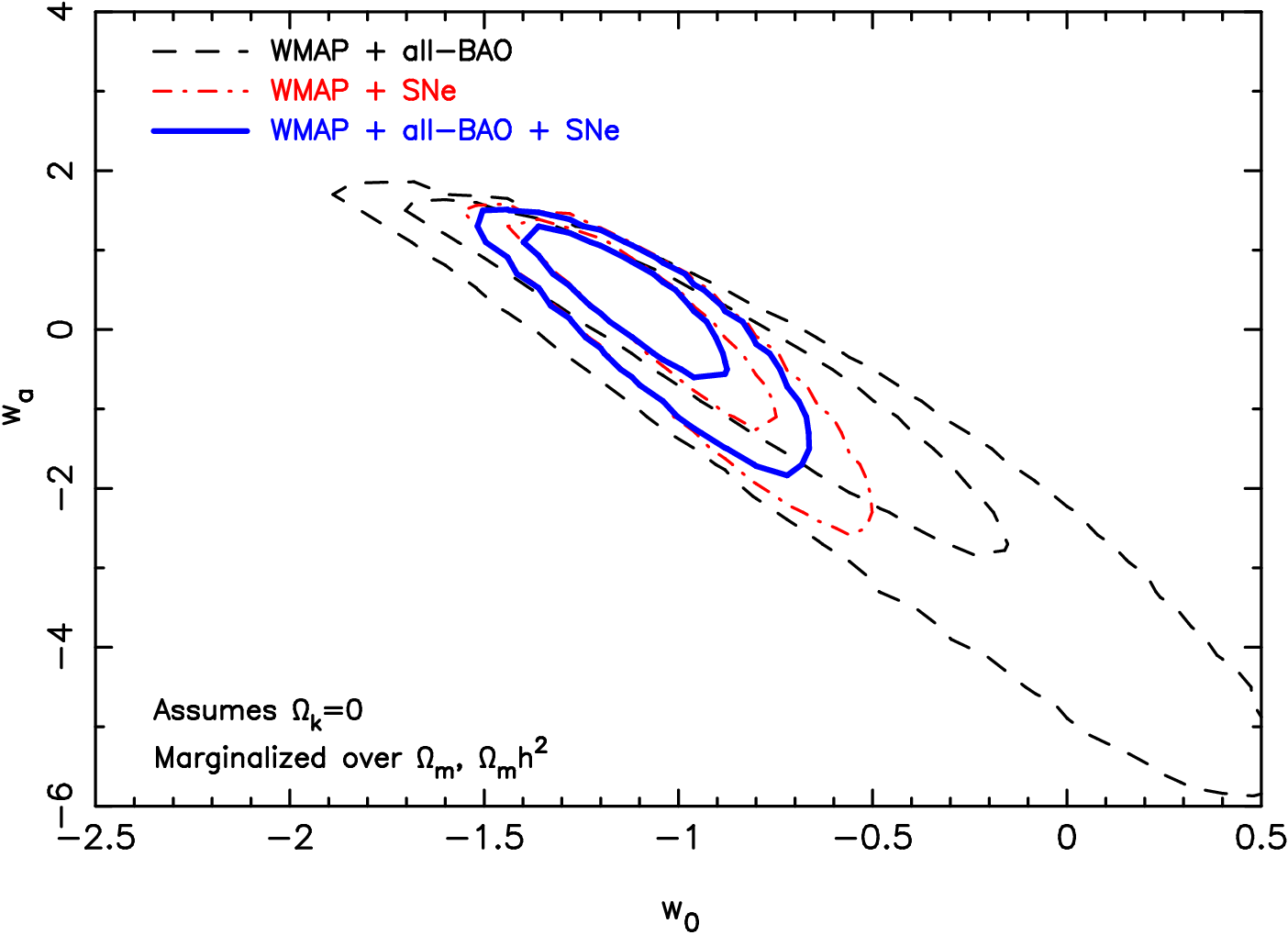}}}
\end{center}
\caption{The joint probability for parameters $w_0$ and $w_a$
  describing an evolving equation-of-state for dark energy, fitted to
  various combinations of WMAP, BAO and SNe distance data,
  marginalized over $\Omega_{\rm m}$ and $\Omega_{\rm m} h^2$ and
  assuming $\Omega_{\rm k} = 0$.  The two contour levels in each case
  enclose regions containing $68.27\%$ and $95.45\%$ of the total
  likelihood.}
\label{figprobw0wa}
\end{figure}

\begin{table*}
\begin{center}
\caption{The results of fitting various cosmological models to a
  combination of the latest CMB, BAO and SNe distance datasets.
  Measurements and 1-$\sigma$ errors are listed for each parameter,
  marginalizing over the other parameters of the model.  All models
  contain either $(\Omega_{\rm m}, \Omega_{\rm m} h^2)$ or
  $(\Omega_{\rm m}, h)$ amongst the parameters fitted.}
\label{tabcosmofit}
\begin{tabular}{ccccccccc}
\hline
Model & $\chi^2$ & d.o.f. & $\Omega_{\rm m}$ & $\Omega_{\rm m} h^2$ & $h$ & $\Omega_{\rm k}$ & $w_0$ & $w_a$ \\
\hline
Flat $\Lambda$CDM & $533.1$ & $564$ & $0.290 \pm 0.014$ & $0.1382 \pm 0.0029$ & $0.690 \pm 0.009$ & - & - & - \\
Flat $w$CDM & $532.9$ & $563$ & $0.289 \pm 0.015$ & $0.1395 \pm 0.0043$ & $0.696 \pm 0.017$ & - & $-1.034 \pm 0.080$ & - \\
Curved $\Lambda$CDM & $532.7$ & $563$ & $0.292 \pm 0.014$ & $0.1354 \pm 0.0054$ & $0.681 \pm 0.017$ & $-0.0040 \pm 0.0062$ & - & - \\
Curved $w$CDM & $531.9$ & $562$ & $0.289 \pm 0.015$ & $0.1361 \pm 0.0055$ & $0.687 \pm 0.019$ & $-0.0061 \pm 0.0070$ & $-1.063 \pm 0.094$ & - \\
Flat $w(a)$CDM & $531.9$ & $562$ & $0.288 \pm 0.016$ & $0.1386 \pm 0.0053$ & $0.695 \pm 0.017$ & - & $-1.094 \pm 0.171$ & $0.194 \pm 0.687$ \\
\hline
\end{tabular}
\end{center}
\end{table*}

\section{Conclusions}
\label{secconc}

We summarize the results of our study as follows:

\begin{itemize}

\item The final dataset of the WiggleZ Dark Energy Survey allows the
  imprint of the baryon acoustic peak to be detected in the galaxy
  correlation function for independent redshift slices of width
  $\Delta z = 0.4$.  A simple quasi-linear acoustic peak model
  provides a good fit to the correlation functions over a range of
  separations $10 < s < 180 \, h^{-1}$ Mpc.  The resulting
  distance-scale measurements are determined by both the acoustic peak
  position and the overall shape of the clustering pattern, such that
  the whole correlation function is being used as a standard ruler.
  As such, the acoustic parameter $A(z)$ introduced by Eisenstein et
  al.\ (2005) represents the most appropriate distilled parameter for
  quantifying the WiggleZ BAO measurements, and we present in Table
  \ref{tabwigcov} a $3 \times 3$ covariance matrix describing the
  determination of $A(z)$ from WiggleZ data at the three redshifts
  $z=0.44$, $0.6$ and $0.73$.  We test for systematics in this
  measurement by varying the fitting range and implementation of the
  quasi-linear model, and also by repeating our fits for a dark matter
  halo subset of the Gigaparsec WiggleZ simulation.  In no case do we
  find evidence for significant systematic error.

\item We present a new measurement of the baryon acoustic feature in
  the correlation function of the Sloan Digital Sky Survey Luminous
  Red Galaxy (SDSS-LRG) sample, finding that the feature is detected
  within a subset spanning the redshift range $0.16 < z < 0.44$ with a
  statistical significance of $3.4$-$\sigma$.  We derive a measurement
  of the distilled parameter $d_{z=0.314} = 0.1239 \pm 0.0033$ that is
  consistent with previous analyses of the LRG power spectrum.

\item We combine the galaxy correlation functions measured from the
  WiggleZ, 6-degree Field Galaxy Survey and SDSS-LRG samples.  Each of
  these datasets shows independent evidence for the baryon acoustic
  peak, and the combined correlation function contains a BAO detection
  with a statistical significance of $4.9$-$\sigma$ relative to a
  zero-baryon model with no peak.

\item We fit cosmological models to the combined 6dFGS, SDSS and
  WiggleZ BAO dataset, now comprising six distance-redshift data
  points, and compare the results to similar fits to the latest
  compilation of supernovae (SNe) and Cosmic Microwave Background
  (CMB) data.  The BAO and SNe datasets produce consistent
  measurements of the equation-of-state $w$ of dark energy, when
  separately combined with the CMB, providing a powerful check for
  systematic errors in either of these distance probes.  Combining all
  datasets, we determine $w = -1.034 \pm 0.080$ for a flat Universe,
  and $\Omega_{\rm k} = -0.0040 \pm 0.0062$ for a curved,
  cosmological-constant Universe.

\item Adding extra degrees of freedom always produces best-fitting
  parameters consistent with a cosmological constant dark-energy model
  within a spatially-flat Universe.  Varying both curvature and $w$,
  we find marginalized errors $w = -1.063 \pm 0.094$ and $\Omega_{\rm
    k} = -0.0061 \pm 0.0070$.  For a dark-energy model evolving with
  scale factor $a$ such that $w(a) = w_0 + (1-a) w_a$, we find that
  $w_0 = -1.09 \pm 0.17$ and $w_a = 0.19 \pm 0.69$.

\end{itemize}

In conclusion, we have presented and analyzed the most comprehensive
baryon acoustic oscillation dataset assembled to date.  Results from
the WiggleZ Dark Energy Survey have allowed us to extend this dataset
up to redshift $z=0.73$, thereby spanning the whole redshift range for
which dark energy is hypothesized to govern the cosmic expansion
history.  By fitting cosmological models to this dataset we have
established that a flat $\Lambda$CDM cosmological model continues to
provide a good and self-consistent description of CMB, BAO and SNe
data.  In particular, the BAO and SNe yield consistent measurements of
the distance-redshift relation across the common redshift interval
probed.  Our results serve as a baseline for the analysis of future
CMB datasets provided by the {\it Planck} satellite (Ade et al.\ 2011)
and BAO measurements from the Baryon Oscillation Spectroscopic Survey
(Eisenstein et al.\ 2011).

\section*{Acknowledgments}

We thank the anonymous referee for careful and constructive comments
that improved this study.

We acknowledge financial support from the Australian Research Council
through Discovery Project grants DP0772084 and DP1093738 funding the
positions of SB, DP, MP, GP and TMD.  SC and DC acknowledge the
support of the Australian Research Council through QEII Fellowships.
MJD thanks the Gregg Thompson Dark Energy Travel Fund for financial
support.

We thank the LasDamas project for making their mock catalogues
publicly available.  In particular EK is much obliged to Cameron
McBride for supplying mock catalogues on demand.  EK also thanks Ariel
S{\'a}nchez for fruitful lengthy discussions.  EK was partially
supported by a Google Research Award and NASA Award.

FB is supported by the Australian Government through the International
Postgraduate Research Scholarship (IPRS) and by scholarships from
ICRAR and the AAO.

GALEX (the Galaxy Evolution Explorer) is a NASA Small Explorer,
launched in April 2003.  We gratefully acknowledge NASA's support for
construction, operation and science analysis for the GALEX mission,
developed in co-operation with the Centre National d'Etudes Spatiales
of France and the Korean Ministry of Science and Technology.

Finally, the WiggleZ survey would not be possible without the
dedicated work of the staff of the Australian Astronomical Observatory
in the development and support of the AAOmega spectrograph, and the
running of the AAT.

\end{document}